\documentclass[%
 preprint,showpacs,showkeys,
 amsmath,amssymb,euscript,
 aps,
]{revtex4-1}

\usepackage{graphicx}
\usepackage{epstopdf}
\usepackage{dcolumn}
\usepackage{bm}
\usepackage{hyperref} 

\begin{document}

\preprint{APS/123-QED}

\title{Thermodynamics of a one-dimensional self-gravitating gas with periodic boundary conditions}
\author{Pankaj Kumar}
\author{Bruce N. Miller}%
 \email{b.miller@tcu.edu}
\affiliation{Department of Physics and Astronomy,\\
 Texas Christian University, Fort Worth, Texas 76129, USA}
\author{Dan Pirjol}
\affiliation{Department of Particle Physics, National Institute
of Physics and Nuclear Technology, Bucharest, Romania}

\date{\today}

\begin{abstract}
We study the thermodynamical properties of a one-dimensional gas with
one-dimensional gravitational interactions, and placed in a uniform mass
background.
Periodic boundary conditions are implemented as a modification of the
potential consisting of a sum over mirror images (Ewald sum), regularized
with an exponential cut-off. The system has a phase transition at a
critical temperature. Above the critical temperature the gas density is uniform,
while below the critical point the system becomes inhomogeneous.
Numerical simulations of the model confirms the existence of the phase
transition, and are in good agreement with the theoretical results.
\end{abstract}

\pacs{05.70.Fh, 04.40.-b, 05.10.-a, 05.45.Pq}
\keywords{Statistical mechanics; phase transition; $N$-body simulation; Lyapunov exponent} 
\maketitle

\section{\label{sec:1}Introduction\protect\\}
The thermodynamics of self-gravitating systems has been studied for a long time,
starting with the classical analysis of Chandrasekhar \cite{Chandra}.
An overview of the stability problem for such systems in three dimensions
is given by Chavanis in \cite{Chavanis}. The stability properties are found
to be different in the canonical and microcanonical ensembles. In the
canonical ensemble a self-gravitating system enclosed in a volume of radius
$R$ is unstable under collapse at energies below
a critical value $E_c = - 0.335 GM^2/R$, the so-called Antonov instability.

The problem is simplified by working in a one-dimensional setting.
The equilibrium thermodynamics of the one-dimensional gravitational gas has been
studied for a bewildering variety of scalings and assumptions. We start by
giving a brief summary of the results in the literature. For simplicity we
restrict this summary to studies of the one-dimensional gravitational gas in
thermal equilibrium. We also mention only studies where the equation of state
of the gas  is derived from first principles, as opposed to being one of the
inputs of the analysis.

Salzberg \cite{Salzberg} considered a one-dimensional gravitational gas
of $N$ particles of mass $m$ enclosed in a finite volume $L$, and
interacting by potentials $gm^2 |x_i-x_j|$ with a hard core $d$. (Note that one-dimensional particles correspond to
three-dimensional mass sheets which can move freely and cross each other).
This leads to non-extensive thermodynamics in the limit
$N \to \infty$ at fixed $m$. For example the total interaction
energy scales like $U \simeq N^3$ as $N\to \infty$.
The equation of state has the form $L = Nd + 2kT/p$, which is
essentially the free gas equation of state corrected by the hard core
volume $p=kT/(L - Nd)$. This is clearly not very realistic, so alternative
scalings for the interaction $gm^2$ with $N$ have been explored in
the literature.

A different setting was adopted by Rybicki \cite{Rybicki}, who considered
$N$ particles of mass $m$ moving along the infinite line and interacting
by potentials $V(x,y) = gm^2 |x-y|$ (no hard core).
The $N\to \infty$ limit was taken at fixed total mass $M=Nm$ and
total energy $E$ (Vlasov limit). This corresponds to scaling the particle
masses as $m = M/N$.
The one-particle distribution function was computed, from which the
density of the gas in thermodynamical equilibrium was obtained.
Under the infinite volume setup assumed in \cite{Rybicki}, the equation
of state of the gravitational gas was not considered in this paper.

Considering a gas enclosed into a finite volume $[0,L]$, the usual
thermodynamical limit is $N, L\to \infty$ at fixed particle number density
$\rho=N/L$. This limit was considered by Isihara \cite{Isihara} who studied
the equilibrium thermodynamics of a one-dimensional gas enclosed in the
box $[0,L]$ interacting with two-body potentials
\begin{eqnarray}\label{VIsihara}
V(x,y) = \left\{
\begin{array}{cc}
- \frac{\mu}{L} \left(1 - \frac{1}{L} |x-y|\right) & \,, |x-y| > \delta \\
+ \infty & \,, |x-y| < \delta \\
\end{array}
\right.
\end{eqnarray}
Apart from the constant term, this interaction is identical to the
one-dimensional gravitational interaction with strength
$2\pi G m^2$, under the scaling $m \sim 1/L$ for the particle masses
which corresponds to fixed total particle mass $M = mN = m \rho L$.
The advantage of this scaling is that it gives usual extensive properties
for the gas energy and entropy.

The paper \cite{Isihara} derived the thermodynamical quantities of the gas
with interaction (\ref{VIsihara}) under certain special
periodic boundary conditions, and concluded that the equation of state is
van der Waals. The system has a liquid-gas phase transition. This is somewhat
surprising, considering that no such phase transition is obtained for the
one-dimensional gravitational gas in \cite{Rybicki}. However, these systems 
differ in one important respect, as the interaction
(\ref{VIsihara}) has a hard core. In a wide class of interacting systems 
(systems of particles interacting by Kac potentials),
a hard core is required in order to have a phase transition \cite{Presutti}. 

In order to study further this issue, a lattice gas version of the system 
considered in \cite{Isihara} was
studied in \cite{PS}. This can be shown to be equivalent to a continuous
one-dimensional gas enclosed in the box $[0,1]$ with the interaction
\begin{eqnarray}\label{Vxi}
V(x,y)  =  |x-y| -  \xi
\end{eqnarray}
and with a special form of the entropy function, specific to the lattice gas.
The constant $\xi$ is an universal attractive interaction, which is felt by
all particle pairs.
Taking $\xi=1$ reproduces the Isihara interaction (\ref{VIsihara}), 
and taking $\xi=0$ reproduces the
interaction potential of the one-dimensional gravitational gas.
The main result of \cite{PS} is that the system has a phase transition only
for $\xi>0$, while for $\xi=0$ no such phenomenon is observed. The exact
equation of state is obtained in the thermodynamical limit, which turns out
to be different from a van der Waals equation, although it is
qualitatively similar, and it approaches van der Waals form in the large
temperature limit.

A scaling similar to that described above
was proposed by de Vega and Sanch\'ez
\cite{deVegaSanchez} in the context of {\em three-dimensional} systems by taking
the thermodynamical limit $N,R\to \infty$ at fixed $N/R$, with $R$ the size of
the system. This is similar to the one-dimensional scaling considered above.

The equilibrium thermodynamics of the one-dimensional gravitational gas was
also studied by Monahan \cite{Monaghan}, and by Fukui, Morita \cite{FM}.
The paper \cite{Monaghan} derived an
exact lower bound on the partition function of the one-dimensional
gravitational gas following from the Jensen inequality. As shown in
\cite{PS}, such a bound gives an accurate approximation which approaches
the exact result in the large temperature limit.

Periodic boundary conditions are often used in practice to simplify the
solution of statistical mechanics problems. With short-range interactions
they can be shown to preserve the thermodynamical properties of the
system, up to a surface term which has a subleading contribution in the
thermodynamical limit \cite{FL}. While the gravitational
interaction is long-ranged and does not satisfy the conditions under which the
results of \cite{FL} are obtained, modifications of the one-dimensional
interaction with periodic boundary conditions have been considered as well.

One of the best known models of this type in the literature is perhaps the
HMF model with Hamiltonian \cite{HMF}
\begin{eqnarray}
H_{\rm HMF} = \sum_{i=1}^N \frac12 mv_i^2 + \frac{\gamma}{2N}\sum_{i<j}
[1 - \cos(\theta_i - \theta_j)],
\end{eqnarray}
where $\theta_i \in (0,2\pi)$. This corresponds to particles moving on a circle
of unit radius and interacting by attractive potentials $V_{ij} = \frac{\gamma}{4N}
d_{ij}^2$ where the distance between the particles is
$d_{ij} = 2\sin(\frac12(\theta_i-\theta_j))$. Note that the potential
is quadratic in the distance, as opposed to linear as appropriate for
one-dimensional gravity. A similar model is the self-gravitating ring model,
where the particles are constrained to move on a circle, and interact by
three-dimensional gravitational potentials \cite{SGR}.

We consider in this paper an alternative way of introducing periodic boundary
conditions in the one-dimensional gravitational system, which was
proposed by Miller and Rouet in \cite{MR}, and has the advantage of preserving
the linear dependence of the potential on the distance $d=|x-y|$ between
particles for sufficiently small $d$. This corresponds to the following
set-up: the system is enclosed in the box $[0,L]$ and has periodic boundary
conditions. In addition, the system is assumed to be placed
into the uniform background of a mass distribution. The model is appropriate
for studying one-dimensional density fluctuations in a uniform mass distribution and a Coulombic version of the model has been used to investigate single-component plasmas \cite{KM}.

The Miller-Rouet model is somewhat similar to the OSC model (one dimensional static cosmology)
which was introduced by Aurell et al \cite{A1,A2, A3} and studied further
by Valageas in \cite{Val,Val2}. This model differs from the former in how the periodic boundary conditions are implemented.
Specifically, the periodicity is imposed by adding an external potential.
In contrast, the Miller-Rouet model considered in this paper maintains the translation
invariance and implements the periodic condition by modifying only the two-body
interaction potential.

We use classical statistical methods to derive the thermodynamical properties
of the system in the canonical ensemble. We determine the single particle
distribution function by minimizing the free energy, and obtain explicit
results for the free energy density and the equation of state of the gas.
The system is homogeneous for temperatures larger than a critical temperature,
and develops a position-dependent density below this temperature. The states
with inhomogeneous density are states of thermodynamical equilibrium.

MD simulations are generally employed in the study of the systems that
exhibit considerable chaotic dynamics needed to attain a phase-space
equilibrium. A smaller finite-sized version of an otherwise ergodic-like
system may have a segmented phase space with KAM tori separating the
stable and unstable regions. For example, in a three-body version of
the Miller-Rouet gravitational gas, it was shown that the phase-space
always exhibits chaotic as well as stable regions and a KAM breakdown
to complete chaos is not observed at any energy \cite{Kumar2016}.
However, as the number of particles ($N$) is increased, the contribution
from chaotic orbits increases drastically and any randomly selected initial
condition results in a chaotic orbit with LCEs converging to a single
universal value for a given energy \cite{Kumar2016_2}. Such behavior has also
been shown for the free-boundary version of the one-dimensional
gravitational gas system in which the phase space was found to be
practically fully chaotic for $N \geq 5$ \cite{Benettin1979}. In general,
caution must be taken while applying theoretical and MD methods to systems
with segmented phase space.

Phase-space mixing leading to a relaxed state is a prerequisite for
equilibrium statistical mechanics to apply to a system. Mixing in phase
space arises as a result of dynamical instability in the phase space and
is usually characterized by existence of at least one positive Lyapunov
characteristic exponent (LCE) \cite{Krylov1979}. For a Hamiltonian system
with $n$ degrees of freedom, a full Lyapunov spectrum may have up to $n-1$
positive LCEs \cite{Pesin1976,Pesin1977}. Of particular interest is the
maximal positive LCE $\lambda_1$ which quantifies the largest average rate
of exponential divergence of a given phase-space orbit with respect to the
nearby orbits \cite{Pesin1976,Benettin1979,ott2002,sprott2003}.

Apart from being important from a dynamical perspective, LCEs also play an
important role in thermodynamic studies and have been shown to serve as
indicators of phase transitions \cite{Butera1987,Caiani1997,Casetti2000,Dellago1996,Barre2001}. For example, the largest LCE was shown to attain a maximum at the fluid-solid phase transition for a two-dimensional particle system \cite{Dellago1996}. The largest LCE has also been observed to display a transition-like variation at the critical temperature for a one-dimensional chain of coupled nonlinear oscillators \cite{Barre2001}. In $N$-body simulations with a finite number of particles, LCEs have also been shown to exhibit behaviors that are observed in the thermodynamic limit \cite{Bonasera1995,Latora1999}.

An exact numerical method of calculating the full Lyapunov spectrum was proposed for the case of one-dimensional gravitation gas \cite{Benettin1979} and the approach was further extended to the periodic-boundary versions of Coulombic and gravitational versions \cite{Kumar2016_2}. As we shall see in Sec. \ref{sec:Lyap_exponent_160} of this paper, we use the formulations presented in Ref. \cite{Kumar2016_2} to calculate the largest LCE and examine its temperature dependence for an indication of a phase transition.

The paper is structured as follows. For ease of reference we give an overview
of the Miller-Rouet model and of its derivation in Section~\ref{sec:2}.
In Secs. \ref{sec:3} and \ref{sec:4}, we formulate the statistical mechanics for
the model in the canonical ensemble and derive its thermodynamical properties.
We work in the Vlasov limit, by taking the particle number
very large $N\to \infty$ at fixed total particles mass $M=Nm$.
This leads to finite results for the energy and free energy per particle.
The single particle distribution function, giving the gas density, is obtained
by solving a variational problem for the free energy.
In Sec. \ref{sec:5}, we verify the
validity of the theoretical predictions by numerically computing the
time-averaged values of such thermodynamic parameters as temperature,
radial distribution function, and pressure in $N$-body simulations of the
model using a molecular-dynamics (MD) approach. A few technical derivations
are given in two Appendices.

\section{The Miller-Rouet model}
\label{sec:2}

\begin{figure}[t]
    \centering
   \includegraphics[width=3in]{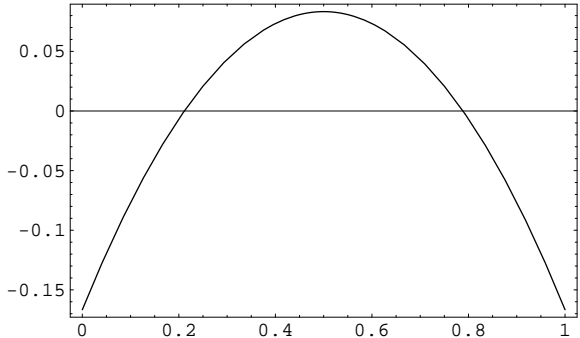}
    \caption{
Plot of the interaction energy $L V(x,y)$ vs $\frac{1}{L}|x-y|$
in the Miller, Rouet model.}
\label{Fig:Vbarxy}
 \end{figure}

We consider in this paper a one-dimensional gas of particles of mass $m$
enclosed in a box $[0,L]$, and interacting with potential energy \cite{MR}
\begin{eqnarray}\label{Vbarsol}
V(x,y) = 2\pi G m^2
\left( |y-x| - \frac{1}{L} (y-x)^2 - \frac{1}{6}L \right) \,.
\end{eqnarray}
This is the potential energy of a mass at position $x$ due to the interaction
with another particle at $y$ and all its mirror images separated by the
periodicity length $L$.
The plot of the potential $V(x,y)$ is shown in Figure~\ref{Fig:Vbarxy}.
At small distances $|x-y| \ll L$ it grows approximatively linearly, just as
the one-dimensional gravitational potential, but for $|x-y| > L/2$ it
becomes repulsive.

We recall briefly the derivation of this potential and its relation to
one-dimensional gravitation. The potential $V(x,y)$ is the difference
of two terms: the sum of the contributions from mirror images $V_0(x,y)$, 
and the contribution of the uniform background of mass $\Phi(x)$
\begin{eqnarray}
V(x,y) = V_0(x,y) - 2\pi G m^2 \frac{1}{L}
  \int_{-\infty}^\infty dy |x-y| e^{-\kappa|x-y|}\,.
\end{eqnarray}

The interaction $V_0(x,y)$ gives the potential felt by a particle placed at
$y$ from a particle at $x$ plus the infinite number of its mirror images,
separated by $L$ in both directions
\begin{eqnarray}\label{sum}
V_0(x,y) =
\sum_{k = -\infty}^\infty 2\pi Gm^2 |x - y + k L| e^{-\kappa |x-y+kL|} \,.
\end{eqnarray}
The damping factor $e^{-\kappa |x-y+kL|}$ with $\kappa \to 0$
is introduced following \cite{Kiessling} and has the advantage that it makes
the sum over mirror images convergent.

The sum over mirror images can be evaluated in closed form with the result
\begin{eqnarray}\label{Vexact}
&& \sum_{k = -\infty}^\infty |x - y + k L| e^{-\kappa |x-y+kL|} \\
&& = |x-y| e^{-\kappa |x-y|} + 2L \frac{e^{\kappa L}}{(e^{\kappa L}-1)^2}
\cosh[\kappa(y-x)]
- 2 \frac{1}{e^{\kappa L}-1} (y-x) \sinh[\kappa(y-x)] \nonumber
\end{eqnarray}
where the first term is the contribution from the $n=0$ term in the sum,
and the remaining terms are the contributions from the mirror images of
the particle at $x$. The proof of this result is given in the Appendix A.

Expanding (\ref{Vexact}) in the limit $\kappa \to 0$
and keeping only the terms which do not vanish in this limit we get
\begin{eqnarray}\label{Vsol}
\lim_{\kappa\to 0} V_0(x,y) = 2\pi G m^2 \left(
|y-x| - \frac{1}{L} (y-x)^2 + \frac{2}{\kappa^2 L} - \frac{1}{6}L \right)\,.
\end{eqnarray}
The first term is the original linear attractive interaction, and the
second term is a quadratic repulsive interaction, which vanishes
in the limit of a very large periodicity radius $L\to \infty$. The physical
meaning of this repulsive term is as follows. As two particles are separated by
more than $L/2$, the attractive effect of their mirror images in the nearby
cells overcomes the
attractive interaction between them. This appears as a repulsive force when
the distance satisfies $|x-y| > L/2$.

Finally we subtract the contribution of a uniform background of mass.
This amounts to a interaction energy $\Phi(x)$ given by
\begin{eqnarray}\label{Phisol}
\Phi(x) = \frac{2\pi Gm^2}{L}
\int_{-\infty}^\infty dy |y-x| e^{-\kappa |x-y|}
= \frac{4\pi Gm^2}{\kappa^2 L} \,.
\end{eqnarray}
This is a uniform potential, independent of position.
The effect of subtracting the uniform background
contribution $\Phi(x)$ from (\ref{Vsol}) amounts to canceling out the
(positive) constant term $4\pi Gm^2/(\kappa^2 L)$.
The remaining constant term $-\frac16 2\pi Gm^2 L$ is negative and finite.


\section{Thermodynamics in the static Vlasov limit}
\label{sec:3}

We would like to derive the thermodynamical properties of a continuous
gas enclosed in the $[0,L]$ volume, with periodic boundary condition,
and interacting by the potential (\ref{Vbarsol}).

We work at fixed $L$ and total particle mass $M = m N$. This implies that
the particle masses scale as $m \sim 1/N$. Since both the inertial and
gravitational mass are scaled, the dynamics of the system is the same as if
the inertial mass is $1$, and the particles interact by the potential
\begin{eqnarray}\label{Vrescaled}
V(x,y) &=& 2\pi G M \frac{1}{N} \left( |x-y| - \frac{1}{L} (x-y)^2 - \frac16 L
\right)\,.
\end{eqnarray}
This requires also that the temperature is rescaled as $T/m \to T$. In order
to make this rescaling explicit, we will denote in this section the rescaled
temperature as $T_V = T/m$
(the subscript stands for Vlasov temperature).

It is well known that a system with interaction of the form (\ref{Vrescaled})
can be described by a mean-field theory
\cite{MS,Bavaud}. The interaction potential scales like $\sim 1/N$ with the number
of particles, at fixed volume $L$. In the limit $N\to \infty$ the energy
per particle approaches a finite value, and the system is described by the
one-particle distribution function $\rho(x)$, giving the probability of
finding a particle in the volume element $[x,x+dx]$.
This limiting procedure corresponds to the mean-field, or static Vlasov limit.

Assume that the system is at a given temperature $T_V$.
The free energy per particle is
\begin{eqnarray}
f = \frac{F}{N} = u - T_V s
\end{eqnarray}
where $f = f_{\rm Q}  + f_{\rm kin}$ consists of a configurational
contribution $f_{\rm Q}$ and a contribution from the kinetic degrees of
freedom $f_{\rm kin} = k_B T_V ( \log N -1 + \frac12 \log(\frac{h^2}{2\pi k_BT_V}))$.
The configurational contribution is given by the solution of the variational
problem
\begin{eqnarray}\label{fQ}
f_{\rm Q} &=& \mbox{inf}_\rho
\left\{ \frac12 (2\pi G M) \int_0^L dx dy \rho(x)\rho(y)
 \left(|x-y| - \frac{1}{L} (x-y)^2 - \frac16 L \right) \right. \\
& & \left. \qquad \qquad + k_B T_V \int_0^L dx \rho(x) \log \rho(x) \right\} \nonumber
\end{eqnarray}
where the infimum is taken over all functions $\rho(x)$ normalized as
\begin{eqnarray}\label{norm}
\int_0^L dx \rho(x) = 1\,.
\end{eqnarray}
The solution of the variational problem (\ref{fQ}) gives the single particle
distribution function $\rho(x)$. This gives the so-called isothermal
Lane-Emden equation for $\rho(x)$ \cite{MS,Bavaud}.

The energy $u=\frac{U}{N}$ and entropy $s=\frac{S}{N}$ per particle are
easily obtained from the free energy $f$ as $u = f - T_V \partial_{T_V} f$ and
$s = - \partial_{T_V} f$. They are given by
\begin{eqnarray}\label{usol}
u &=& \frac12 k_B T_V +
\frac12 (2\pi GM) \int_0^L dx dy \rho(x)\rho(y)
 \left(|x-y| - \frac{1}{L} (x-y)^2 - \frac16 L \right)\\
\label{Sdef0}
s &=& - k_B \int_0^L dx \rho(x) \log \rho(x) + s_{\rm kin}
\end{eqnarray}
where $\rho(x)$ is the minimizer of the functional (\ref{fQ}).
The total energy per particle is the sum of the contribution from the kinetic
energy, and the interaction energy with the remaining $N-1$ particles.
The contribution of the kinetic degrees of freedom to the entropy per
particle is $s_{\rm kin} = - \partial_{T_V} f_{\rm kin} = -k_B (\log N -  \frac32
+ \frac12 \log\frac{h^2}{2\pi k_B T_V} )$.

\subsection{Large temperature approximation}

For temperatures $k_B T_V\gg 2\pi G M \frac14 L$ much larger than the range
of variation
of the potential $V(x,y)$, the density of the gas approaches a constant
value $\rho(x) = 1/L$. Expressed in terms of the actual temperature $T = m T_V$
this condition reads $k_B T \gg \frac14 2\pi G \frac{M^2}{N}$.

In this regime the thermodynamical properties of the gas
simplify very much. The energy and entropy per particle become
\begin{eqnarray}\label{U0}
u &=& \frac12 k_B T_V +
\frac12 g^2  \int_0^L dx dy (|x-y|-(x-y)^2-\frac16) = \frac12 k_B  T_V \\
s &=&  k_B \log (L/N) +
 k_B \left(\frac32 - \frac12 \log\frac{h^2}{2\pi k_B T_V} \right)\,.
\end{eqnarray}
Note that the constant term $-\frac16$ in the interaction energy cancels the
contributions from the linear and quadratic terms, and the total interaction
energy of the gas vanishes in the uniform density limit. The only contribution
in (\ref{U0}) comes from the kinetic degrees of freedom.

The total free energy of the gas is
\begin{eqnarray}\label{fsol}
F &=& N(u - T_V s) = \frac12 N k_B T_V - N k_B T_V \log (L/N) - N k_B T_V
\left(\frac32 - \frac12 \log\frac{h^2}{2\pi k_B T} \right) \\
 &=& L \left( \frac12 k_B \bar\rho  T_V
+ k_B T_V  \bar\rho \log\bar\rho + k_B T_V \bar\rho \log \bar\rho \right)
  - L \bar\rho k_B T_V
\left(\frac32 - \frac12 \log\frac{h^2}{2\pi k_B T_V} \right)  \,. \nonumber
\end{eqnarray}
In the last line we introduced $\bar\rho = N/L$ the particle number density
of the gas.

The pressure of the gas is
\begin{eqnarray}\label{eqstate}
p(\bar\rho,T_V) = - (\partial_L F)_{N,T_V} =  k_B \bar\rho T_V \,,
\end{eqnarray}
which is the ideal gas equation of state.

\subsection{An energy-entropy argument}

For temperatures $T_V$ comparable to $2\pi G M \frac14 L$ and below, the
thermodynamics of the system with interaction (\ref{Vbarsol}) is expected to
be more complex. We give next a qualitative discussion based on an energy-entropy
minimization argument. The equilibrium state is given in general by the
minimum of the free energy $F=U-TS$. For $T=0$ this corresponds to the minimum
of $U$, while for $T\to \infty$ it corresponds to a maximum of the entropy $S$.

For small temperatures $T\to 0$ the equilibrium state of the system
corresponds to a minimum of the total energy. An examination of the plot
of the interaction energy $V(x,y)$ in Figure~\ref{Fig:Vbarxy} shows that
the system has two possible ground states: i) a state where the particles
are grouped together into one block (minimal separation), and ii) a state
where the particles are
separated into clumps separated by a distance $L$.
These two states correspond to the minima of the interaction potential $V(x,y)$,
see Figure~\ref{Fig:Vbarxy}.

On the other hand, in the infinite temperature limit, the equilibrium state
corresponds to the maximum of the entropy, which is given by the uniform density
state studied in the previous section. This state is unique. Therefore, as the
temperature is lowered, we expect that at some critical temperature we have
a bifurcation (or transition) where the system condenses into one of the two
ground states, or into a combination of them.

The situation is very similar to that encountered in the OSC model \cite{Val}
which is also a model of one-dimensional gravitation,
in a uniform background of mass, and with periodic boundary conditions.
The treatment of the periodic boundary conditions is different, and results
in a non-trivial external potential $\Phi(x)$.  The Hamiltonian of this model is
\begin{eqnarray}
H_{\rm OSC} = \sum_{i=1}^N \frac12 m v_i^2 + gm^2 \sum_{i>j} |x_i-x_j| -
gm\bar \rho \sum_{i=1}^N \left[\left(x_i-\frac12 L\right)^2 + \frac14 L^2\right]\,.
\end{eqnarray}
Each particle feels the potential interaction with the uniform background
$\Phi(x) = - gm\bar\rho [(x-\frac12 L)^2 + \frac14 L^2]$ which has the
effect of pushing the particles towards the ends of the box $x\to 0$ and
$x\to L$. The combined effect of the linear attraction potential, and of
the external potential $\Phi(x)$ is to produce a complex phase diagram, with
several phase transitions.

\section{Solution of the model}
\label{sec:4}

We derive in this Section the exact result for the thermodynamical properties
of the system in the canonical ensemble, for arbitrary temperature. First we
simplify the problem by taking
without any loss of generality the size of the box to be $L=1$. The parameter
$L$ can be absorbed into a redefinition of the coordinate $x/L \to x$.
Second, for notational simplicity we denote the rescaled temperature $T_V = T/m$ simply
as $T$.  We will convert back to $T$ in Section 5, in order to compare the
theoretical predictions with the numerical simulation.

\subsection{Lane-Emden equation}

The single particle distribution function is found by solving the Lane-Emden
equation. The result is given by the following Proposition.

{\bf Proposition 1.}
{\em The single particle distribution function of the gas
in thermodynamical equilibrium $\rho(x)$ with
$x\in [0,1]$ satisfies the Lane-Emden equation}
\begin{eqnarray}\label{LE}
\frac{d^2}{dx^2} \log\rho(x) = 2\beta (1 - \rho(x))
\end{eqnarray}
{\em normalized as}
\begin{eqnarray}\label{const}
\int_0^1 dx \rho(x) = 1 \,.
\end{eqnarray}

{\bf Proof.} The functional $f_Q[\rho]$ for the configurational contribution
to the free energy per particle (\ref{fQ}) can be written as
\begin{eqnarray}\label{fQnorm}
f_{\rm Q}[\rho] &=& \mbox{inf}_\rho
\left\{ \frac12 g^2 \int_0^1 dx dy \rho(x)\rho(y)
 \left(|x-y| - (x-y)^2 - \frac16  \right) \right. \\
& &  \qquad \qquad \left.
+T \int_0^1 dx \rho(x) \log \rho(x) - T \log L \right\} \nonumber
\end{eqnarray}
where we introduced $g^2 = 2 \pi G ML$. The function $\rho(x)$ appearing in
this expression is a rescaled density and is related as $\rho(x) = L \tilde \rho(xL)$
where $\tilde \rho(y)$ is the density appearing in (\ref{fQ}).
For simplicity we assume in the remainder of the paper
that the Boltzmann constant is $k_B=1$.
At equilibrium the free energy is minimal. We would like to minimize
$F$ under the constraint (\ref{const}). This
constraint can be taken into account by introducing a Lagrange
multiplier $\lambda$ and considering the functional $G[\rho] =
f_Q[\rho] +\lambda (\int_0^1 dx \rho(x) - 1)$.

This variational problem gives the Euler-Lagrange equation
for the gas density $\rho(x)$.
\begin{eqnarray}\label{EL}
\frac{\delta}{\delta\rho(x)} G[\rho] = g^2
\int_0^1 dy \rho(y) \left( |x-y| - (x-y)^2 - \frac16 \right) +
T (\log\rho(x) + 1) +\lambda = 0
\end{eqnarray}
This integral equation can be transformed into a differential equation
by taking 2 derivatives with respect to $x$. Writing explicitly the first
integral, the Euler-Lagrange equation is written as
\begin{eqnarray}
&& g^2\left(
\int_0^x dy \rho(y)(x-y) + \int_x^1 dy \rho(y) (y-x) - \int_0^1 dy\rho(y)
(x-y)^2 - \frac16 \right) \\
&& \qquad + T (\log\rho(x) + 1) + \lambda = 0\,. \nonumber
\end{eqnarray}
Taking one derivative with respect to $x$ we get
\begin{eqnarray}
g^2 \left(
\int_0^x dy \rho(y) - \int_x^1 dy \rho(y) -2 \int_0^1 dy \rho(y)(x-y)\right)
+ T \frac{d}{dx} \log\rho(x) = 0\,.
\end{eqnarray}
Take a second derivative
\begin{eqnarray}
2g^2 \rho(x) - 2g^2 + T \frac{d^2}{dx^2} \log\rho(x) = 0
\end{eqnarray}
This is the Lane-Emden equation (\ref{LE}), which holds for the single particle
distribution function for a gas at temperature $T$ \cite{MS}.
This concludes the proof of this relation.

We would like to solve the equation (\ref{LE}) with the constraint
(\ref{const}), for given temperature $T$. It is convenient to introduce the
new unknown function $y(x)$ defined by
\begin{eqnarray}
\rho(x) =  e^{y(x)}
\end{eqnarray}
Expressed in terms of this function, the differential equation (\ref{LE}) reads
\begin{eqnarray}\label{LE2}
y''(x) = 2\beta g^2 (1- e^{y(x)})
\end{eqnarray}
with the normalization constraint
\begin{eqnarray}\label{const2}
\int_0^1 dx e^{y(x)} = 1\,.
\end{eqnarray}
We impose periodic boundary conditions
\begin{eqnarray}\label{BC}
y(0) = y(1) \,,\qquad y'(0) = y'(1) \,.
\end{eqnarray}

We note that the equation (\ref{LE2}) is identical to equation (12)
in \cite{Val} (up to the redefinition $y(x) \to - \beta \psi(x)$ and rescaling
$x/L \to x$), giving the density of the gas in the OSC model.
However our boundary conditions (\ref{BC}) are more constraining than the
boundary condition in \cite{Val}. In particular, we require $y(0)=y(1)$,
which is not imposed in \cite{Val}.
As a result, although the qualitative properties of the
solution are similar in both cases, the details of the solution are different.

{\bf Remark 1.} {\em We note that the normalization constraint (\ref{const})
is automatically satisfied with the boundary condition $y'(0) = y'(1)$.
Indeed, using the equation (\ref{LE2}) we have}
\begin{eqnarray}
\int_0^1 dx e^{y(x)} = \int_0^1 dx \left( 1 - \frac{1}{2\beta g^2 } y''(x)
\right) = 1 - \frac{1}{2\beta g^2 } (y'(1)-y'(0)) = 1\,.
\end{eqnarray}

We write the equation (\ref{LE2}) in the form
\begin{eqnarray}\label{yeqmotion}
y''(x) = - V'(y(x))\,, \quad V(y) = \alpha^2 (e^y - y -1)
\end{eqnarray}
where we defined $\alpha^2 = 2\beta g^2$. This has same form as the
Newton's equation of motion for a particle of mass 1 in the potential
$V(y)$. The total energy is conserved
\begin{eqnarray}
E = \frac12 [y'(x)]^2 + \alpha^2 (e^{y(x)} - y(x)) \,.
\end{eqnarray}

Using this dynamical analogy it is easy to understand  the qualitative
behavior of the solutions of the differential equation (\ref{yeqmotion}).
The equation (\ref{yeqmotion}) has always the trivial solution $y(x)=0$,
which corresponds to the particle sitting at rest at the bottom of the potential $V(y)$. In addition to this trivial solution it can have oscillatory solutions,
corresponding to the particle moving in the potential $V(y)$, starting
at some non-zero value $y(0)\neq 0$ with a positive or negative initial speed
$y'(0)$, and then performing one full oscillation or several oscillations before
returning to the starting point $y(1)=y(0)$ with the same velocity $y'(1)=y'(0)$
at time $1$. The movement of the particle is spanned by $y_L \leq y(x) \leq y_R$,
where $y_L<0, y_R>0$ are the turning points at which the particle speed vanishes.
They are related by energy conservation to the initial position and speed as
$V(y_L) = \frac12 [y'(0)]^2 + V(y(0)) = V(y_R)$.
It is easy to see that one can take $y'(0)=0$ without any loss of generality,
as the solutions with non-zero $y'(0)$ are related to those with $y'(0)=0$ by
a translation.

We will be seeking solutions of the equation (\ref{yeqmotion}) with
boundary conditions $y(0)=y(1), y'(0)=y'(1)=0$ corresponding to the particle
starting at rest at time 0 at $y(0)$ and returning to the same position
at time 1. There are two solutions which are distinguished by the sign of the
initial position: $y^{(+)}(0) > 0$ and $y^{(-)}(0) > 0$. However, it is easy
to see that they are related by a translation $x\to x+C$, and it is sufficient
to determine only one of them. We will choose as the representative solution
the solution with $y(0)<0$.  From this one can generate a continuous family
of solutions by translations in  the $x$ coordinate.

\begin{figure}
\centering
\includegraphics[width=4in]{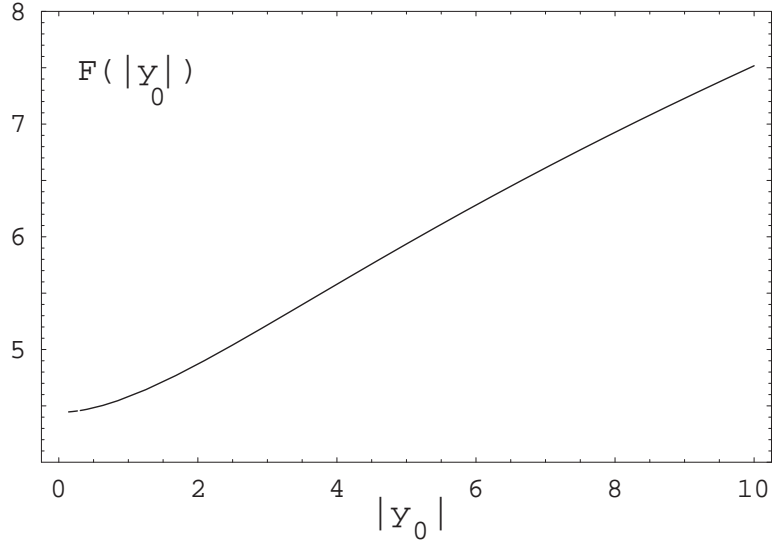}
\caption{
Plot of the function $F(y_0)$ defined in (\ref{Fy0}) for $y_0<0$ vs $|y_0|$.}
\label{Fig:3}
\end{figure}

The solution $y(x)$ is given implicitly by
\begin{eqnarray}\label{eqyx}
x = \int_{y(0)}^{y(x)} \frac{dy}{\sqrt{2\alpha^2 (y - e^y - y(0) + e^{y(0)})}}
\end{eqnarray}
where $y(0)$ is a solution of the equation
\begin{eqnarray}\label{y0eq}
2k t(y_0, \alpha) = 1\,,\qquad k = 1, 2, \cdots
\end{eqnarray}
$t(y_0,\alpha)$ is the time it takes the particle to move from $y_0<0$
when starting at rest $y'(0)=0$, to the turning point with opposite sign
$y_R(y_0)$.
This function is given by
\begin{eqnarray}\label{Fy0}
t(y_0, \alpha) = \frac{1}{\sqrt{2\alpha^2}} \int_{y_0}^{y_{R}(y_0)}
\frac{dy}{\sqrt{y - e^y - y_0 + e^{y_0}}} \equiv \frac{1}{\sqrt{2}\alpha}
F(y_0)\,,
\end{eqnarray}
with $y_R(y_0)>0$ the positive solution of the equation $e^{y_0}-y_0=
e^{y_R(y_0)}-y_R(y_0)$.
We defined the function $F(y_0)$ as the integral appearing in this expression.
The plot of $F(|y_0|)$ is shown in Figure~\ref{Fig:3}.
It has the limiting value $\lim_{y_0\to 0} F(y_0)=\sqrt{2}\pi$, and it is an
increasing function of $|y_0|$.

The solutions of the equation (\ref{y0eq}) with $k=1$ describe trajectories
where the particle performs one full oscillation before returning to $y(0)$
at $x=1$, the solutions with $k=2$ give trajectories with two oscillations and
so on. Equation (\ref{y0eq}) has both positive and negative solutions for $y_0$.
As discussed above, it is sufficient to consider only the $y_0<0$ solution.
We will denote the solution corresponding to given $k$ as $y_k(x)$ and will
call it the $k-th$ mode.

It is clear that the equation (\ref{y0eq}) has solutions for given
$k\in \mathbb{N}$ only if $\frac{2\pi k}{\alpha} > 1$. In particular,
for $\alpha < 2\pi$ this equation does not have a non-zero solution for $y_0$,
and the only solution of the equation of motion
(\ref{yeqmotion}) is the trivial solution $y(x)=0$. For $2\pi < \alpha < 4\pi$
there is solution $k=1$,
for $4\pi < \alpha < 6\pi$ there are two solutions with $k=1,2$, and so on.
We give in Table~\ref{Table:1} a tabulation of the $k=1$ solutions of the
equation (\ref{y0eq}) for values of $\alpha>2\pi$.

The higher order solutions are related to the $k=1$ solution as
\begin{eqnarray}\label{ykx}
y_1(x,\alpha) = y_2\left(\frac12 x,2\alpha\right) = \cdots =
y_k\left(\frac{1}{k}x,k\alpha\right)\,.
\end{eqnarray}
It is easy to check by direct substitution into the equation $y''(x,\alpha)
= \alpha^2(1-e^{y(x,\alpha)})$ that these are indeed solution of this equation.
In particular, this gives a relation among the solutions of the equation
(\ref{y0eq}) with different values of $k$: $y_1(0,\alpha) = y_2(0,2\alpha) =
\dots$.

For sufficiently small oscillation amplitude $|y_0| \ll 1$ the function
$t(y_0)$ is given by the approximative formula
\begin{eqnarray}
t(y_0) = \frac{\pi}{\alpha} \left(1 + \frac{1}{24} y_0^2 + O(y_0^4) \right)\,.
\end{eqnarray}
This follows from the expansion of the oscillation period for an anharmonic
potential with small amplitude, see \cite{osc} for a detailed discussion and
references to the literature. The small amplitude region $|y_0| \ll 1$ corresponds
to $\alpha$ just above $2\pi$ (for $k=1$), just above $4\pi$ (for $k=2$), etc.

Using this approximation we get the solution of (\ref{y0eq}) for values of
$\alpha$ around the critical value $2\pi$
\begin{eqnarray}\label{y0app}
y_0^2 \simeq \left\{
\begin{array}{cc}
0 & \,, \alpha \leq 2\pi  \\
24\left( \frac{\alpha}{2\pi } -1 \right) & \,, \alpha > 2\pi  \\
\end{array}
\right.
\end{eqnarray}
A similar formula gives $y_k(0)$ for $\alpha$ slightly above $2\pi k$ with
$k=2,3,\cdots$. This is $y_k^2(0) \simeq 24(\frac{\alpha}{2\pi k}-1)$.

For $|y_0| \ll 1$ we can find also an explicit approximative
solution of the equation (\ref{eqyx}). This is given for $k=1$ by
\begin{eqnarray}\label{cosapp}
y_1(x) = y_1(0) \cos(2\pi x)
\end{eqnarray}
This follows by expanding the exponentials in the denominator of the integrand
in a Taylor series to second order, which gives
\begin{eqnarray}
x = \frac{1}{\alpha} \int_{y_0}^{y(x)} \frac{dy}{\sqrt{y_0^2-y^2}} =
\arccos\left(\frac{y(x)}{y_0}\right)\,.
\end{eqnarray}
The properly normalized density of the gas is 
$\rho(x) = \frac{1}{I_0(y_1(0))} e^{y_1(0) \cos(2\pi x)}$.

Combining (\ref{cosapp}) with (\ref{ykx}) one can obtain also approximative
solutions for $y_k(x)$ for $\alpha$ slightly larger than $2\pi k$.

The analysis presented above gives the following qualitative behaviour of the
gas density as the temperature is lowered. In the infinite temperature limit
$T\to \infty$ we have $\alpha\to 0$ and the gas density is constant
$\rho(x) = 1$.
As the temperature is lowered, the density $\rho(x)$ remains constant until
we reach $\alpha = 2\pi$ when one non-trivial solution for $y_0$ appears.
This point corresponds to temperature
\begin{eqnarray}\label{Tc1}
T_{c1} = \frac{g^2}{2\pi^2} = \frac{2\pi G ML}{2\pi^2}\,.
\end{eqnarray}

Compared to the critical temperature in the OSC model \cite{Val},
this is smaller by a factor of $\frac14$. This is due to our boundary condition
$y(0)=y(1)$ which is not imposed in \cite{Val}.
However, the result for $T_{c1}$ has the same dependence on model
parameters as in the OSC model, see Eq.~(11) in \cite{Val2} which gives
$T_{c1} = \frac{2g^2}{\pi^2}$ in our notations.
Note that in this reference $2\pi G$ is denoted $g$.

\begin{figure}[t]
    \centering
   \includegraphics[width=2.5in]{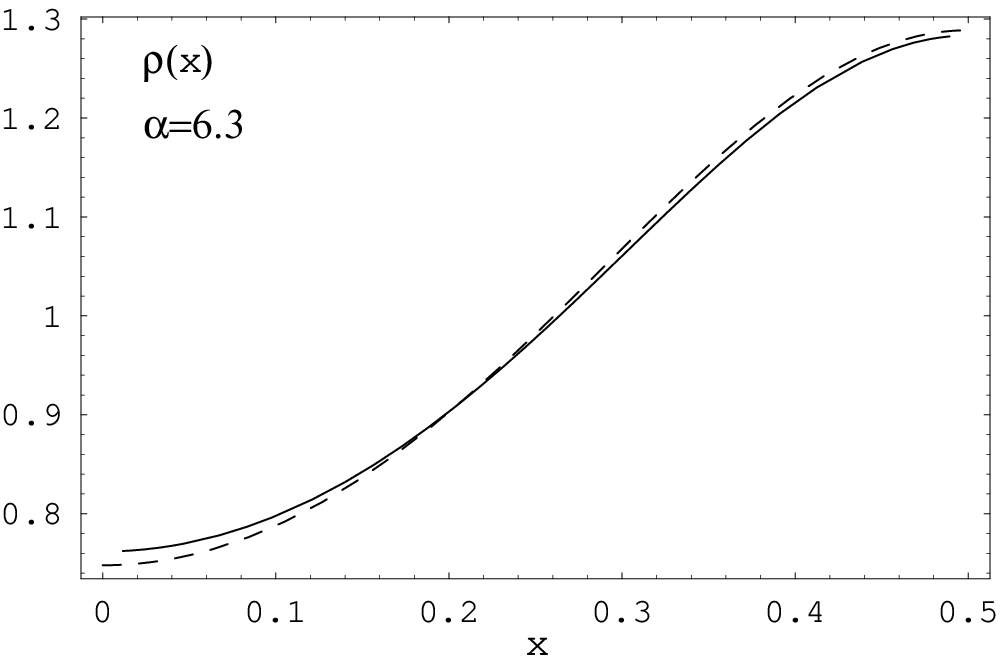}
   \includegraphics[width=2.5in]{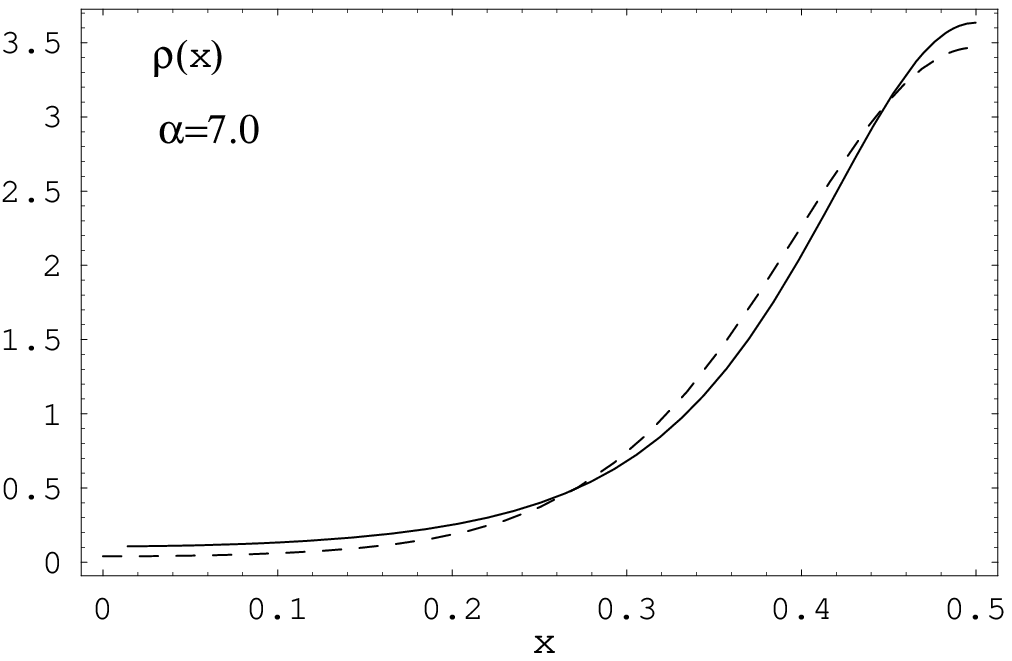}
    \caption{
Plots of the density $\rho(x)$ vs $x$ for several values of
$\alpha^2=2\beta g^2$ (solid curves). The dashed curves show the
approximation (\ref{cosapp}) which is valid for $\alpha \sim 2\pi$.
Left: $\alpha = 6.3$, right: $\alpha=7$. }
\label{Fig:2}
 \end{figure}

As the temperature is lowered below this point, non-trivial solutions
with inhomogeneous gas density appear. They are translated versions of
the basic solution $\rho_1(x) = \exp(y_1(x))$.
$\rho_1(x)$  has a maximum at $x=1/2$. We show in Figure~\ref{Fig:2}
typical results for the gas density profiles $\rho_1(x)$ for
two values of $\alpha = 6.3$ (just above $2\pi$) and $\alpha = 7$ (solid curves). These are
compared with the approximation (\ref{cosapp}) (dashed curves) which is
seen to work well for temperature just below the transition temperature
$T_{c1}$.

As the temperature is lowered further, we reach the point $\alpha = 4\pi$,
corresponding to temperature $T_{c2} = \frac{g^2 ML}{8\pi^2}$. Below
this temperature there are two solutions for $y(x)$ corresponding to $k=1,2$.
In addition to the $k=1$ solution we have another
solution with $k=2$, which has oscillatory density behavior, and has two
maxima/minima within the box. In general there is an infinite
sequence of critical temperatures at which new solutions appear, given by
$\alpha = 2n\pi$, with $n=1,2,\cdots$
\begin{eqnarray}
T_{cn} = \frac{g^2 }{2n^2 \pi^2 } = \frac{2\pi G ML}{2n^2 \pi^2}\,.
\end{eqnarray}

Note that we have not yet proven that these solutions of the Lane-Emden equation
for the gas density $\rho(x)$ correspond to stable
configurations of the gas. In order to decide which solutions are stable
one has to compare their free energy and determine the solution which minimizes
the free energy. This will be done in the next section.

\subsection{Thermodynamics}

We discuss next the thermodynamical properties of the system. They can be
obtained from the free energy $F$, which is given by the following result.

{\bf Proposition 2.} {\em The free energy per particle of the gas is given by}
\begin{eqnarray}\label{Fexact}
\frac{F}{N} = \frac{1}{2g^2} T^2 \int_0^1 dx [y'(x)]^2 +
T (1 - e^{y_0} + y_0 ) - T\log L + f_{\rm kin}\,,
\end{eqnarray}
{\em where $y_0<0$ is the solution of the equation $\sqrt{\frac{2}{\alpha}} F(y_0)=1$
with $F(y_0)$ defined in (\ref{Fy0}). The contribution from the kinetic degrees
of freedom is $f_{\rm kin} = T(\log N - 1 + \frac12 \log\frac{h^2}{2\pi T} )$.
The integral in the first term depends only on $\alpha^2=2g^2\beta$
and is given explicitly for the $k-$th solution of the equation (\ref{y0eq}) as}
\begin{eqnarray}\label{Kdef}
K_k(\alpha) \equiv \int_0^1 dx [y_k^\prime(x)]^2 =
2k \int_{y_0}^{y_{R}(y_0)}
dx \sqrt{2\alpha^2 (x - e^x - y_0 + e^{y_0}) }
\end{eqnarray}
{\em Recall that $y_R(y_0)>0$ is the unique positive
solution of the equation $e^{y_0}-y_0 = e^{y_R(y_0)}-y_R(y_0)$
with $y_0<0$. This is the turning point in
the equivalent dynamical interpretation of the equation satisfied by $y(x)$.}

{\bf Proof.} See Appendix B.

We note that in the $T\to \infty$ limit, we have $\alpha \to 0$ and the
free energy (\ref{Fexact}) reduces to the free energy of the uniform gas
which is given in (\ref{fsol}).

For $\alpha$ slightly above $2\pi$ (corresponding to temperature $T$ just below
the first critical point $T_{c1}$), we can derive a closed form approximation
for the function $K_1(\alpha)$ by expanding the exponential function in the
integrand. This gives
\begin{eqnarray}\label{Kapp}
K_1(\alpha) = 2\alpha y_0^2 \int_{-1}^{1} dx  \sqrt{1-x^2} = \pi
\alpha y_0^2\,.
\end{eqnarray}

We study the behavior of the free energy around the critical
point $T_{c1}$.  The free energy per particle is
\begin{eqnarray}
f = f_0 + \frac{1}{2g^2} T^2 K_1(\alpha) + T(1-e^{y_0} + y_0)
\end{eqnarray}
where $f_0 = T (\log(N/L) -1 + \frac12 \log\frac{h^2}{2\pi T})$ is the free
energy per particle in the homogeneous density phase, below the critical
temperature.

Substituting here the approximations for $K_1(\alpha)$ and $y_0^2$
(\ref{Kapp}) and (\ref{y0app}) we have, for $\alpha$ just above $2\pi$
\begin{eqnarray}
f - f_0 &\simeq& \frac{1}{2g^2} T^2 \pi \alpha y_0^2 + T (-\frac12 y_0^2) \\
&=& \frac12 T y_0^2 \left(\frac{2\pi}{\alpha} - 1\right) =
- \frac{24\pi T}{\alpha} \left( \frac{\alpha}{2\pi} - 1 \right)^2 \,. \nonumber
\end{eqnarray}
Expressing $\alpha$ in terms of temperature as
$\frac{\alpha}{2\pi} = \sqrt{\frac{T_{c1}}{T}}$ we have
\begin{eqnarray}
f - f_0 \simeq - 12 T_{c1} \sqrt{x}(1-\sqrt{x})^2
\end{eqnarray}
with $x \equiv T/T_{c1}$.
It is easy to see that the free energy difference $f-f_0$ and its first
derivative with respect to temperature vanish at $T=T_{c1}$, while the
second derivative has a jump from 0 at $T >T_{c1}$ to
$\lim_{T\to T_{c1}-0} \partial_T^2 (f-f_0) = -6 T_{c1}$. Since the difference 
$f-f_0$ vanishes
for $T>T_{c1}$, this implies that the free energy and its derivative are
continuous at $T=T_{c1}$ while its second derivative has a jump.
We conclude that the
phase transition at $T=T_{c1}$ is a second order phase transition.

We study further the properties of the system around  the critical temperature
$T_{c1}$. The energy per particle of the gas is given by (\ref{usol}).
This can be written in a more explicit way as
\begin{eqnarray}\label{Uexact}
u &=& \frac12 T - \frac{1}{4g^2} T^2 \int_0^1 dx [y'(x)]^2 \,.
\end{eqnarray}

The first term is the kinetic energy contribution, which is given by the
equipartition theorem as $\frac12k_B T$ per particle. The second term is
the contribution of the interaction energy, which can be expressed in this
form using the Lane-Emden equation as shown in Appendix B
\begin{eqnarray}
\frac12 (2\pi GML) \int_0^1 dx dy \rho(x)\rho(y) (|x-y| - (x-y)^2-\frac16)
= - \frac12 T J - \frac12 (\lambda + T)
\end{eqnarray}
The integral $J=\int_0^1 dx \rho(x)\log\rho(x)$ and Lagrange multiplier
$\lambda$ are given explicitly in Appendix B. Substituting their expressions
here gives the result (\ref{Uexact}).

For temperature above the critical temperature, the gas density is uniform
and the contribution of the interaction energy vanishes. The energy per
particle is due in this region only to the kinetic degrees of freedom.
Below the critical temperature, the gas becomes non-uniform
and the interaction energy starts to contribute a non-vanishing amount.

We compute next the specific heat per particle.
This can be obtained by taking a derivative of (\ref{Uexact}) with
respect to the temperature and is given by
\begin{eqnarray}\label{cVres}
c_V &=& \left(\frac{\partial u}{\partial T}\right)_L = \frac12 + \frac{1}{8\pi}
\sqrt{\frac{T}{T_{c1}}}
K_1'(\alpha) - \frac{1}{4\pi^2} \frac{T}{T_{c1}} K_1(\alpha)
= \frac 12 + 3 F_{c_V}(\alpha) \,,
\end{eqnarray}
where we defined
\begin{eqnarray}\label{FcVdef}
F_{c_V}(\alpha) \equiv \frac{1}{12\alpha} K_1'(\alpha) - \frac{1}{3\alpha^2} K_1(\alpha)
\end{eqnarray}
This was obtained by writing $\alpha = 2\pi \sqrt{T_{c1}/T}$ and using
(\ref{Tc1}). For temperatures just below $T_{c1}$ we can approximate
$K_1(\alpha)$ using (\ref{Kapp}). Using this approximation we have
$\lim_{\alpha \to 2\pi} K_1(\alpha) = 0\,,
\lim_{\alpha \to 2\pi} K'_1(\alpha) = 24\pi$, which gives
\begin{eqnarray}
\lim_{\alpha \to 2\pi} F_{c_V}(\alpha) = 1\,.
\end{eqnarray}
This implies that the specific heat is discontinuous at the critical point.
Above the critical point $T_{c1}$ the specific heat is constant and equal to
$c_V=\frac12$, and below the critical point it takes  the value
\begin{eqnarray}\label{cVlim}
\lim_{T \to T_{c1}-\epsilon} c_V = 3 + \frac12 = \frac72 \,.
\end{eqnarray}

We can obtain an approximation for the temperature
dependence of the specific heat per particle $c_V(T)$ below the
critical temperature, using the approximation (\ref{Kapp}) for the
function $K_1(\alpha)$.
This gives the following approximation for $F_{c_V}(\alpha)$
defined in (\ref{FcVdef}), valid for $\alpha - 2\pi \ll 1$
\begin{eqnarray}
F_{c_V}(\alpha) \simeq \frac{6\pi}{\alpha} - 2\,.
\end{eqnarray}
The corresponding result for the specific heat per particle ia
\begin{eqnarray}\label{cVapp}
c_V(T) \simeq \frac12 + 3(3\sqrt{T/T_{c1}}-2) \,.
\end{eqnarray}

\begin{figure}[t]
    \centering
   \includegraphics[width=4in]{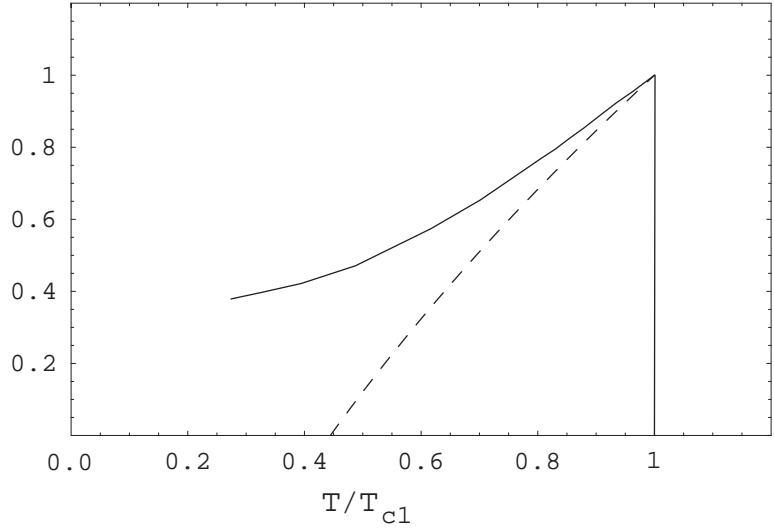}
    \caption{
Plot of the specific heat per particle $\frac13(c_V(T) - \frac12)$ vs $T/T_{c1}$.
This is the excess of the
specific heat over the constant value $\frac12$ which it takes
in the homogeneous phase (times $1/3$). The specific heat has
a finite jump at $T_{c1}$. The solid curve shows the exact result (\ref{cVres})
and the dashed curve shows the approximation (\ref{cVapp}) valid near the
critical point.}
\label{Fig:cV}
 \end{figure}

We show in Figure~\ref{Fig:cV} the plot of the specific heat $c_V(T)$ vs
$T/T_{c1}$. The solid curve is the exact result (\ref{cVres}), and the
dashed curve shows the approximative result (\ref{cVapp}).

\subsection{The higher modes}

The properties of the $k\geq 2$ modes can be related to those of the
$k=1$ mode. This implies that it is sufficient to study the solution of the
system for the $k=1$ mode. This is given by the following relations.

{\bf Proposition 3.}
{\em The solutions of the equation (\ref{y0eq}) are related
as
\begin{eqnarray}\label{id1}
y_k(0,\alpha) = y_1(0,\frac{1}{k}\alpha)
\end{eqnarray}
and the integrals (\ref{Kdef}) are related as}
\begin{eqnarray}\label{id2}
K_k(\alpha) = k^2 K_1(\frac{1}{k}\alpha)
\end{eqnarray}

For simplicity, we prove these relations for $k=2$. The generalization to
arbitrary $k\geq 2$ is immediate.
These relations follow from equation (\ref{ykx}). For $k=2$ this gives
$y_2(x,2\alpha) = y_1(2x,\alpha)$. Taking here $x=0$ gives the first identity
(\ref{id1}).
The second identity (\ref{id2}) is proved as
\begin{eqnarray}
K_2(2\alpha) &=& \int_0^1 dx [y'_2(x,2\alpha)]^2 = 2\int_0^{1/2} dx [y'_2(x,2\alpha)]^2 =
8 \int_0^{1/2} dx [y'_1(2 x,\alpha)]^2 \\
 &=&
4 \int_0^{1} dx [y'_1(x, \alpha)]^2 = 4 K_1(\alpha) \nonumber
\end{eqnarray}

\subsection{Stability analysis}

The difference between the free energy of the $k-$th mode and of the uniform
density state ($k=0$) is obtained by taking the difference of (\ref{Fexact}) and
(\ref{fsol}). This can be written as
\begin{eqnarray}\label{DeltaF}
\Delta f_k \equiv f(y_k(x)) - f_0 =
T \left( 1 - e^{y_0} + y_0 + \frac{1}{\alpha^2} K_k(\alpha)
\right) \equiv  T \delta_k(\alpha)
\end{eqnarray}
where we defined $\delta_k(\alpha)$. This function has a simple interpretation
in terms of the dynamical analogy of the anharmonic oscillator discussed above.

{\bf Remark 4.} {\em The function $\delta_k(\alpha)$ is
related to the classical action along  the trajectory of the equivalent
dynamical system discussed above as}
\begin{eqnarray}
{\cal S}[y(x)] = \int_0^1 dx
\left[ \frac12 [y'(x)]^2 - \alpha^2 (e^{y(x)} - y(x)) \right]\,.
\end{eqnarray}
{\em It is easy to see that we have}
\begin{eqnarray}
\delta_k(\alpha) = 1 + \frac{1}{\alpha^2} {\cal S}[y_k(x)]\,.
\end{eqnarray}

The function $\delta_k(\alpha)$ is tabulated numerically for the first two
modes in Table~\ref{Table:1} in Appendix B. From these results one observes that
$\Delta f_{1,2}$ is negative for temperatures below $T_{c1}$ corresponding
to $\alpha > 2\pi$. We prove next this result analytically for $k=1$.

For temperatures just below the first critical temperature $T_{c1}$ the
free energy difference of the $k=1$ mode is given approximatively by
\begin{eqnarray}
\delta_1 = \frac12 \pi T \alpha y_0^2 + 1 - e^{y_0} + y_0  \simeq
 1 - e^{y_0} + y_0 + \pi T \sqrt{\frac{T}{2\bar\rho}} y_0^2
\end{eqnarray}
where we used the approximation (\ref{Kapp}) which is valid
for $|y_0| \ll 1$, just below the first critical temperature $T_{c1}$.
This is written equivalently as
\begin{eqnarray}
\delta_1 \simeq
1 - e^{y_0} + y_0 + \frac12 T \sqrt{\frac{T}{T_{c_1}}} y_0^2
\leq
1 - e^{y_0} + y_0 + \frac{1}{4\pi^2} y_0^2
\end{eqnarray}
where we used $T < T_{c1}$.

It is easy to see  that the function
$f(x) = 1 - e^x + x + \frac{1}{4\pi^2} x^2$
is strictly negative for any $x\neq 0$. This can be seen either by explicit
numerical evaluation or can be proved analytically as follows.
For $x>0$ it follows from the inequality
$e^x \geq 1 + x + \frac12 x^2$ (which is valid for $x\geq 0$),
and for $x<0$ it follows from the inequality
$e^x \geq 1 + x^2/2$ (which is valid for any real $x$).
We have proven thus that the first mode $y_1(x)$ is a stable
equilibrium state for the gas at temperatures just below the first critical
temperature $T < T_{c1}$.
The gas density becomes inhomogeneous in this region, and
has a unimodal shape with a maximum or minimum at the center of the box.

Next we study the stability of the higher modes. Proposition 4 implies the
following result
\begin{eqnarray}\label{deltaksol}
\delta_k(k\alpha) = \delta_1(\alpha)
\end{eqnarray}
This shows that it is sufficient to compute the free energy of the $k=1$ mode
and we obtain automatically also the free energies of the higher modes.
For example these relations give $\delta_2(2\alpha) = \delta_1(\alpha)$,
which can be checked to hold indeed on the numerical results in
Table~\ref{Table:1}.

Numerical calculation of $\delta_1(\alpha)$ shows that it is a monotonously
decreasing function of $\alpha$, which is zero at $\alpha=2\pi$ and decreases
to larger and larger (in absolute value) negative values as $\alpha$ increases.
The relation (\ref{deltaksol}) implies that the $k=1$ mode has the lowest
free energy at all temperatures below the first critical point $T<T_{c1}$.
The higher modes $k\geq 2$, when they exist (for temperatures below the
corresponding critical temperatures) are unstable minima of the free energy,
and the system will always relax into a $k=1$ state.

\section{Numerical simulation}
\label{sec:5}

We present in this Section the results of a numerical simulation of the model.
The simulation solved numerically the dynamical equations of motion of $N$
particles interacting by the potential (\ref{Vbarsol}). Since the dynamical behavior of the system only depends on the net gravitational field experienced by the gravitating sheets (henceforth referred to as ``particles'' or ``bodies''), we drop the constant term in the potential energy for simplicity. Hence, the potential energy for a system with primitive cell of length $L$ and containing $N$ particles may be expressed as
\begin{equation} \label{eq_potential_energyNP}
V = -2\pi Gm^2 \sum_{j=2}^{N} \sum_{i=1}^{j-1} \left(\frac{(x_j-x_i)^2}{L} - |x_j-x_i|\right),
\end{equation}
where $x_i$ and $x_j$ represent the primitive-cell positions of the $i$-th and the $j$-th particles respectively, with $x \in \left[-L/2,L/2\right)$. It should be noted that Miller and Rouet considered an ``expanding-universe'' version of the gravitational system whereby the positions of the particles were expressed in comoving spatial coordinates. Equations of motion were derived and it was shown that the choice of comoving coordinates invoked a damping factor in the equations of motion. The exact evolutions of each particles' positions and velocities were implicit in the derived equations of motion. However, expressions for the time dependence were not explicitly provided. Here, for the sake of completion, we provide the time-dependencies in fixed (non-comoving) coordinates and discuss the evolution algorithm briefly.\\

Following Ref. \cite{MR} for non-comoving spatial coordinates, it can be shown for an ordered system ($x_1 < x_2 < \hdots < x_N$) that
\begin{equation} \label{eq_relMotionGrav}
\frac{d}{dt} W_j(t) = \frac{d^2}{dt^2}Z_j(t) = 2\pi mG \left\lbrace \frac{2N}{L}Z_j(t) -2\right\rbrace ,
\end{equation}
where $v_j$ is the velocity of the $j$-th particle, $Z_j \equiv (x_{j+1}-x_j)$, and $W_j \equiv (v_{j+1}-v_j)$. Solutions to Eq. (\ref{eq_relMotionGrav}) provide the displacements and velocities of $(j+1)$-th particle with respect to those of $j$-th particle in between events of interparticle crossing:
\begin{equation}
Z_j (t)=\frac{L}{N} + \frac{1}{2} \left\lbrace Z_j (0)-\frac{L}{N} +\frac{W_j (0)}{\Lambda}\right\rbrace e^{{\Lambda}t} +  \frac{1}{2} \left\lbrace Z_j (0)-\frac{L}{N} - \frac{W_j (0)}{\Lambda}\right\rbrace e^{{-\Lambda}t} ,
\end{equation}
\begin{equation}
W_j (t)=\frac{\Lambda}{2} \left\lbrace Z_j (0)-\frac{L}{N} +\frac{W_j (0)}{\Lambda}\right\rbrace e^{{\Lambda}t} -  \frac{\Lambda}{2} \left\lbrace Z_j (0)-\frac{L}{N} - \frac{W_j (0)}{\Lambda}\right\rbrace e^{{-\Lambda}t} ,
\end{equation}
where $\Lambda \equiv \sqrt{\frac{4\pi mGN}{L}}$.

Crossing times, $t_{c_j}$ may be obtained analytically as the smaller positive
root (out of the two possible real ones) of $Z_j (t_{c_j}) = 0$.
We find the crossing times using an event-driven algorithm similar to ones
discussed in Refs.~\cite{MR,KM}. The algorithm keeps track of the
evolution by assigning an identifying label to each particle. Once a crossing
occurs, the algorithm interchanges the labels and the velocities of the two
participating particles at the crossing location. In the following iteration,
the algorithm treats the updated system as a new, ordered one but maintains
the original labels, thereby allowing for correct tracking of each particle's
position and velocity. At the end of each iteration, positions $x_j$ and
velocities $v_j$ are obtained respectively from $Z_j$ and $W_j$ by utilizing
the contraints set forth by the conservation of momentum on the position and
velocity of the center of mass \cite{KM}.

In the simulation, we adopt a set of dimensionless units and rescale the
system parameters as follows: $2\pi G = 0.5$, the total mass per unit cell,
$mN = 1$, and the unit-cell size, $L = 1$. Consequently, the characteristic
frequency of the system, $\Lambda = 1$. Without losing generality, we set the
initial velocity of the center of mass to zero.

With the ability to follow the exact time evolution, we study the thermodynamic
behavior of the system for different values of $N$ and, for each $N$, with
varying per-particle energy. For the system to exhibit ergodic-like behavior,
we avoid low values of $N$ \cite{Kumar2016_2}, i.e., we choose $N\geq 20$.
A molecular-dynamics approach then predicates that the time-averaged values
of the thermodynamic quantities will converge to those in the thermodynamic
limit when $N$ becomes sufficiently large.

\subsection{Kinetic energy}

\begin{figure}[t]
\begin{center}
\includegraphics[scale=1]{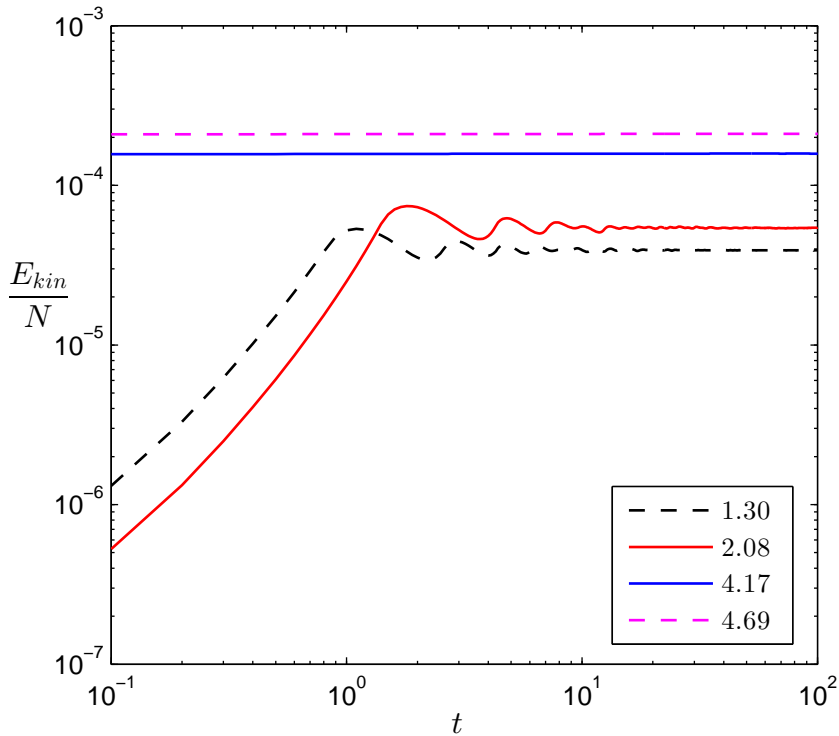}
\caption{\label{fig:KE_convergence} Illustration of convergence of kinetic energy per particle in simulation for the first $100$ time units out of a total time of $1200$. The numbers in the legend represent the value of $\frac{U}{N} (\times 10^{-4})$. }
\end{center}
\end{figure}

\begin{figure}[t]
\begin{center}
\includegraphics[scale=1]{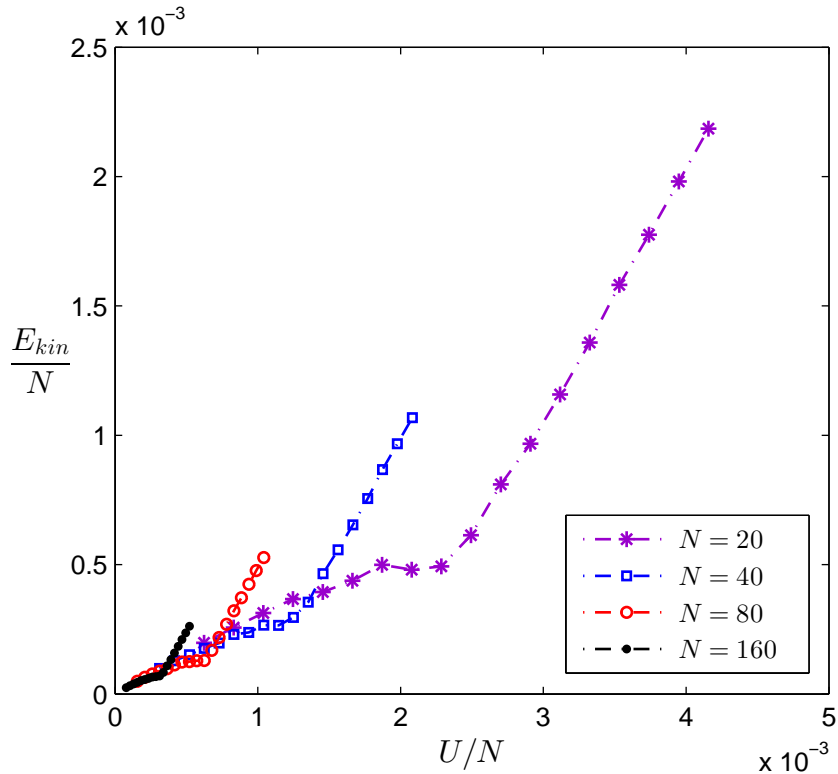}
\caption{\label{fig:VaryingN_KE} Per-particle kinetic energy plotted against per-particle energy for varying number of particles. }
\end{center}
\end{figure}

\begin{figure}[t]
\begin{center}
\includegraphics[scale=1]{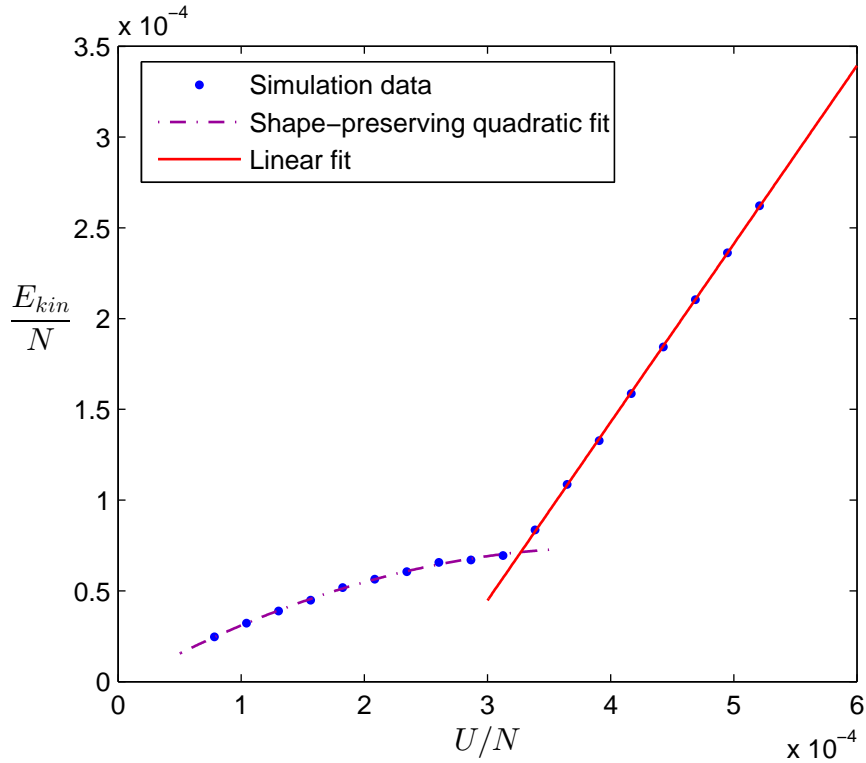}
\caption{\label{fig:N_160_with_fit_curves} Per-particle kinetic energy versus per-particle energy for $N=160$. A clear-cut jump is evident in the first derivative of the caloric curve. }
\end{center}
\end{figure}

\begin{table}[b] \caption{\label{Table:N_160_with_fit_curves}
Converged values of kinetic energy per particle from simulation for $N=160$ found at the four values of per-particle energies used in Fig. \ref{fig:KE_convergence}. In each case, the total simulation time was 1200 and the standard deviation $\sigma_{kin}$ was calculated for the last 200 time units.}
\begin{center}
\begin{tabular}{ccccc}
\hline
$\frac{U}{N} (\times 10^{-4})$ & \hspace{2pt} & $\frac{E_{kin}}{N} (\times 10^{-4})$ & \hspace{2pt} & $\frac{\sigma_{kin}}{(E_{kin}/N)} (\times 10^{-4})$ \\
\hline
1.30 & \hspace{2pt} & 0.39 & \hspace{2pt} & 2.67 \\
2.08 & \hspace{2pt} & 0.56 & \hspace{2pt} & 9.92 \\
4.17 & \hspace{2pt} & 1.59 & \hspace{2pt} & 1.70 \\
4.69 & \hspace{2pt} & 2.10 & \hspace{2pt} & 0.19 \\
\hline
\end{tabular}
\end{center}
\end{table}
Per-particle kinetic energies ($E_{kin}/N$) have been found by sampling the velocities at fixed intervals and averaging the per-particle kinetic energies from each interval over a sufficiently long time. The simulation are first run for $t=1200$, and if the standard deviation $\sigma_{kin}$ from the last $200$ time units have converged to within a set tolerance with respect to the average value, the simulation is terminated. Otherwise, the simulation is allowed to run until the last $200$ time units produce a standard deviation smaller than the tolerance. In our simulations, we specified a tolerance of $1$ percent, that is, $\sigma_{kin} \leq (0.01\times E_{kin}/N)$.

Figure \ref{fig:KE_convergence} shows the evolution of $E_{kin}/N$ for the first $100$ time units out of a total evolution time of $t=1200$ at four different values of per-particle energy $U/N$ for a system with $N=160$. Table \ref{Table:N_160_with_fit_curves} shows the converged value of $E_{kin}/N$ for the same four energies and the corresponding values of $\sigma_{kin}$ relative to $E_{kin}/N$. Evidently, if the value of $U/N$ exceeds the maximum allowed per-particle potential energy, the system acts as an ideal gas and average values of$E_{kin}/N$ converges very rapidly. On the other hand, at energies lower than than maximum allowed values, the system goes through a relaxation phase before the time-averaged values of $E_{kin}/N$ converge.

Figure \ref{fig:VaryingN_KE} shows the caloric curves, $E_{kin}/N$ versus $U/N$, for $N = 20, 40, 80,$ and $160$. Although a transitioning trend is observed for each $N$, the results indicate that the system approaches a ``thermodynamic-limit'' behavior at $N \sim 80$, that is, the transitions become sharp for $N\geq 80$. The caloric curve for $N=160$ has been reproduced separately in Fig. \ref{fig:N_160_with_fit_curves}. A discontinuity in the first derivative is profoundly evident.

\subsection{Radial distribution function} \label{sec_gr}
\begin{figure}[t]
\begin{center}
\includegraphics[scale=0.6]{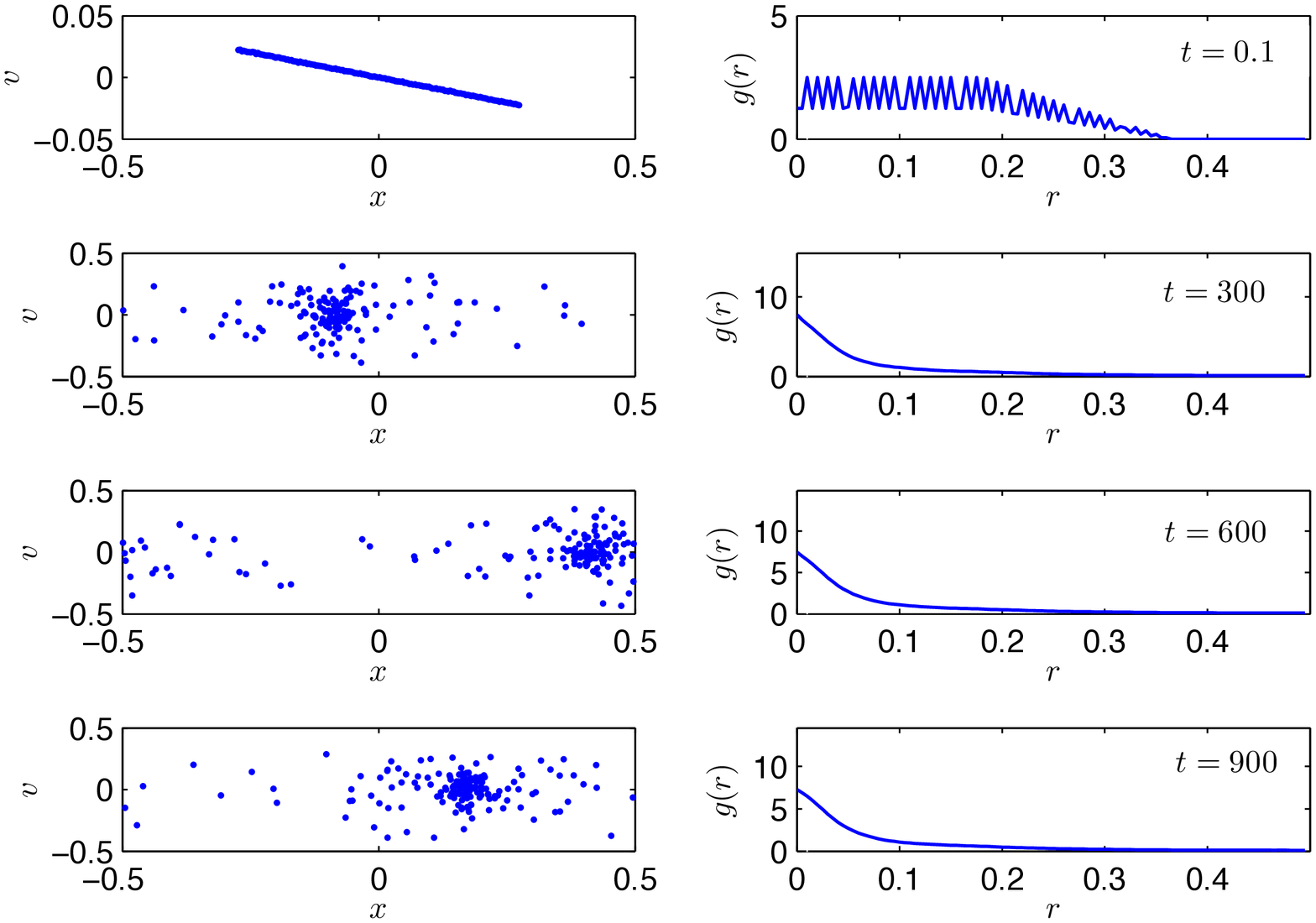}
\caption{\label{fig:g_r_1} $\mu$-space snapshots (left column) and 
time-averaged values of the radial distribution function (right column) for 
$U/N = 2.08 \times 10^{-4}$ and $N=160$ at different instants of time. 
Corresponding elapsed time for each row is mentioned in the $g(r)$ plot.}
\end{center}
\end{figure}

\begin{figure}[t]
\begin{center}
\includegraphics[scale=0.6]{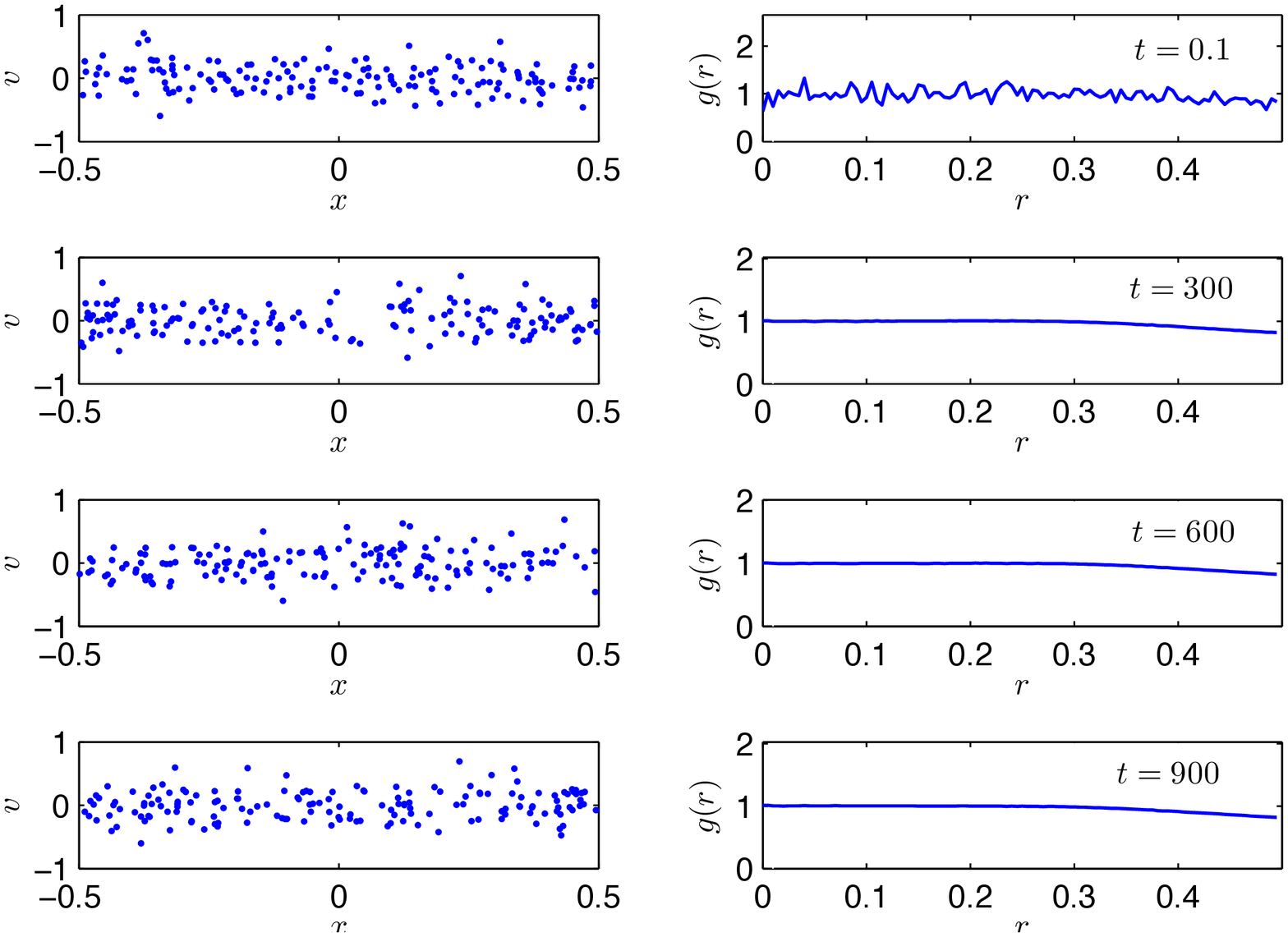}
\caption{\label{fig:g_r_2} $\mu$-space snapshots (left column) and 
time-averaged values of the radial distribution function (right column) for 
$U/N = 4.17 \times 10^{-4}$ and $N=160$ at different instants of time. 
Corresponding elapsed time for each row is mentioned in the $g(r)$ plot. }
\end{center}
\end{figure}

The radial distribution function, $g(r)$ encapsulates how the density varies
with respect to distance $r$ from a reference particle in a system.
To calculate $g(r)$ in simulation, we employ the approach proposed in
Ref. \cite{KM}. The positions of the particles are sampled at fixed
time-intervals of $d\tau$. At the end of $k$-th interval (corresponding to
time, $t = kd\tau$), we find the number of particles, $\Delta N_j(r,t)$ in
a small volume (length) element $\Delta r$ at a distance $r$ from a reference
particle at $x_j$. The radial distribution function is then found as
\begin{equation} \label{eq_correlationFunct}
g(r) = \lim_{l \to \infty} \frac{\sum_{k=0}^l \sum_{j=1}^N \Delta N_j(r,t=kd\tau)}{(2\Delta r)Nl\bar{\rho}} ,
\end{equation}
where $l$ is the number of iterations and $\bar{\rho} = N/L$. Note that,
in Ref. \cite{KM}, the bulk number-density, $\bar{\rho}$ was chosen to be
unity, and hence, it was not included in the expression for $g(r)$.
In our simulation, however, $L$ has been set to unity, and therefore,
$\bar{\rho}$ is simply equal to $N$.

For systems that are homogeneous (and isotropic, in case of two- or three-dimensional systems), $\bar{\rho}g(r)\mbox{d}r$ represents the probability of observing a second particle in $\mbox{d}r$ at a distance $r$ provided a particle is located at $r=0$ and $g(r)\to \frac{N-1}{N}$ for large $r$. \cite{mcquarrie2000}. However, in our case, the system remains essentially non-homogeneous at low energies and the time-averaged value of the density at a position $\textbf{\mbox{r}}$ relative to any given particle at $\textbf{\mbox{r}}_i$ is not equal to the space-averaged bulk density $\bar{\rho}$. That is, $\left\langle \rho(\textbf{\mbox{r}} - \textbf{\mbox{r}}_i) \right\rangle \neq \bar{\rho}$.
Therefore, for low-energy configurations of the Miller-Rouet gravitational gas, $g(r)$ as expressed in Eq. (\ref{eq_correlationFunct}) does not quite represent the standard definition of the radial distribution function as generally used in statistical mechanics. However, it still serves as a good indicator of the relative distribution of the particles with respect to one another. Figure \ref{fig:g_r_1} shows typical low-energy $\mu$-space distributions and the corresponding plots of $g(r)$ at different values of elapsed time. It is evident that the particles tend to stay clumped together and the particle distribution is inhomogeneous.

At high energies, the particles are able to spread across the entire primitive cell and the distribution tends to be homogeneous. That is, $\left\langle \rho(\textbf{\mbox{r}} - \textbf{\mbox{r}}_i) \right\rangle \sim \bar{\rho}$ for $U>V_{max}$, where $V_{max}$ represents the maximum allowed value of the potential energy for a given number of particles. Under such conditions, $g(r)$ as given in Eq. (\ref{eq_correlationFunct}) represents the radial distribution function in the standard sense. Figure \ref{fig:g_r_2} shows a set of high-energy $\mu$-space distributions and the corresponding plots of $g(r)$ at different instants of time. Clearly, the distribution is more uniform in this case (as compared to Fig. \ref{fig:g_r_1}) and $g(r)$ appears to approach the expected value of $\frac{159}{160}$ away from $r=0$.

\subsection{Pressure} \label{sec:Pressure_sim_160}
\begin{figure}[t]
\begin{center}
\includegraphics[scale=1]{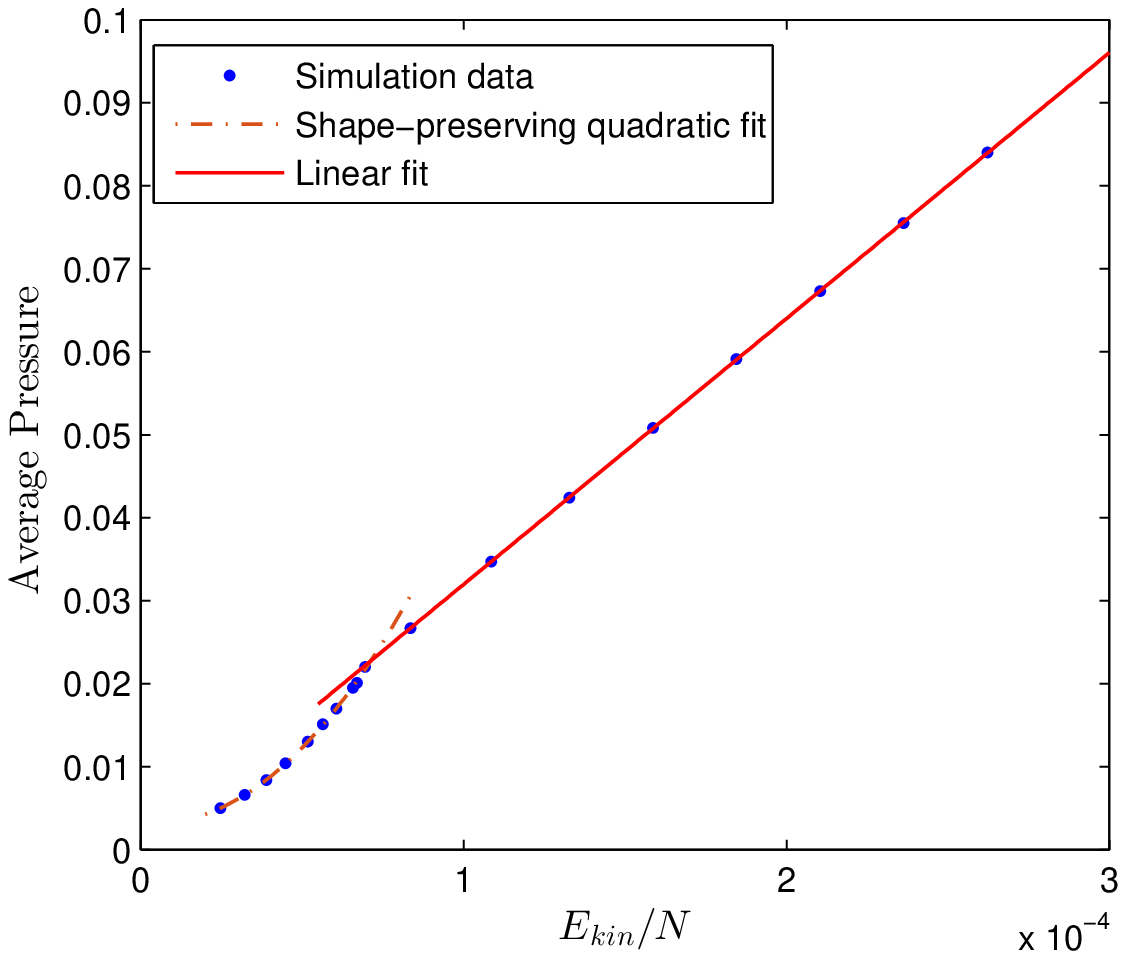}
\caption{\label{fig:N_160_Pressure_fit_curves} Time-averaged pressure
versus per-particle kinetic energy for $N=160$. }
\end{center}
\end{figure}

The pressure has been calculated in simulation by following the method
discussed in Ref. \cite{KM}. The method involves placing virtual walls at
regular spatial intervals throughout the primitive cell and time-averaging
the momentum transferred from hypothetical elastic collisions to each wall
from a given direction (left or right side of the wall). The wall separation
and the averaging time are decided by an adaptive algorithm that takes into
account a user-provided tolerance as the convergence criterion.

Before we start calculating the pressure, the system is allowed to relax
for $t=1200$. Positions and velocities after the initial run of $1200$ time
units are then used as initial conditions for the pressure routine. For
energies greater than the maximum allowed potential energy in the system,
the value of pressure converged fairly easily to within $1$ percent in
$t=500$, with as few as $10$ walls per unit length for $N=160$. The relatively
easy convergence may be attributed to the the fact that the behavior of the
system resembles that of an ideal gas for energies greater than the critical
value of $U$,  However, for energies lower than that corresponding to the
critical point, we had to increase the convergence tolerance to $5$ percent
for the adaptive algorithm to terminate eventually. At a $5$-percent
tolerance, convergence times varied between $t=800$ and $t=1400$ with
$100$ walls for $N=160$ and energies below the critical value.

It should be noted that the particles are tightly coupled via potential at
lower energies and the time evolutions of particles' positions and velocities
are strongly exponential between events of crossings (as opposed to being
uniform, ``ideal-gas-like'' for higher energies). Hence, at energies below
the critical value, finding pressure as an average rate of momentum
transferred by placing regularly-spaced virtual walls becomes a rather
crude approximation. To counter the effect of the strong coupling on the
accuracy of the results, one would have to put increasingly larger number
of virtual walls as the energy gets closer to the critical value. However,
the marginal increase in the accuracy from inserting additional walls
diminishes drastically as the interactions get stronger, thereby making
the simulation increasingly time-consuming for a given convergence tolerance.
Nevertheless, as shown in Fig. \ref{fig:N_160_Pressure_fit_curves}, a
$5$-percent tolerance provided a fairly good handle on the temperature
dependence of pressure for $N=160$, and a clear-cut change in slope is
displayed near the critical value of $E_{kin}/N$.

\subsection{Largest Lyapunov exponent} \label{sec:Lyap_exponent_160}
\begin{figure}[t]
\begin{center}
\includegraphics[scale=1]{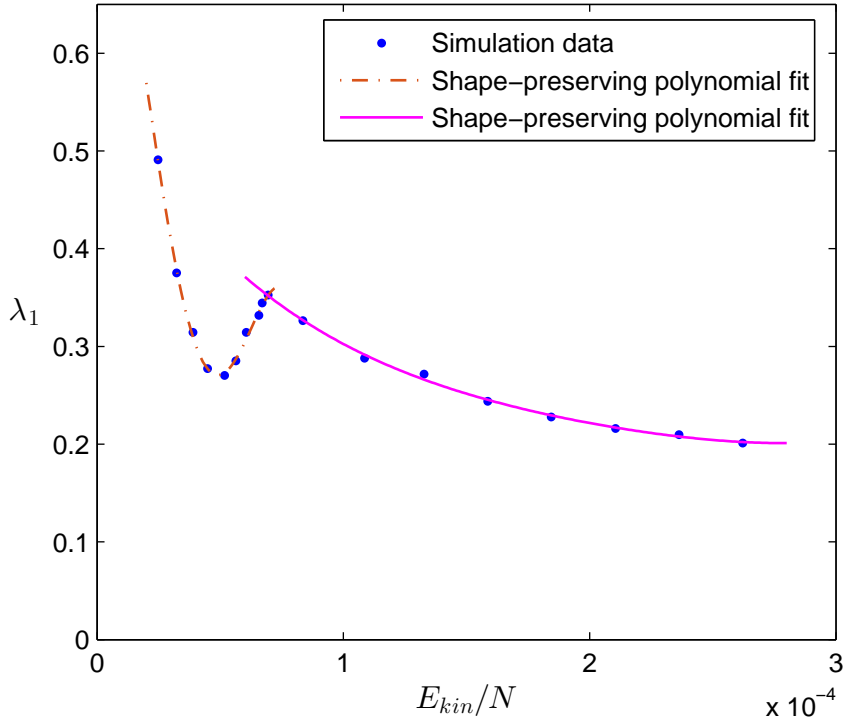}
\caption{\label{fig:N_160_Lyap_fit_curves} Largest Lyapunov characteristic
exponent versus per-particle kinetic energy for $N=160$. }
\end{center}
\end{figure}

The largest Lyapunov characteristic exponents (LCEs) have been calculated for $N=160$ using the method discussed in Ref. \cite{Kumar2016_2}. Similar to the pressure algorithm, the LCE routine uses the positions and velocities from a prior relaxation run of $1200$ time units as the initial conditions. The program is adaptive in that it is allowed to run as long as the standard deviation of the largest Lyapunov exponent from the last $1$ million crossings is greater than $1$ percent of the average value, with a  minimum of 4 million crossings. We found that the largest LCE for each $U/N$ converged to within $1$ percent in the first $4$ million crossings. Results for $N=160$ have been presented in Fig. \ref{fig:N_160_Lyap_fit_curves}. It can be seen from the figure that the largest LCE exhibits a local maximum as well as a discontinuity in the slope near the transition point.

\section{Discussion}
In this section, we compare the results of our simulation with those predicted
by our theoretical treatment. We recall that the theoretical results are
expressed in terms of the Vlasov rescaled temperature $T_V = T/m$.
In order to compare with the simulation, we have to express the theoretical
predictions in terms of the usual temperature $T=T_V m = T_V M/N$.
We also express the energies ($U$ and $E_{kin}$) in the
rescaled units that were adopted in the simulation.

The simulation used the following parameter values: interaction coupling
$2\pi G=\frac12$, total
gas mass $M = mN=1$ and gas volume (length of elementary cell) $L=1$.
Thus we have $g^2 = 2\pi G ML = \frac12$.
Also, the constant term $g^2 (-\frac16 L)$ in the potential (\ref{Vrescaled})
was not included in the simulation, and its effect has to be explicitly subtracted
from the theoretical result.

\subsection{Energy per particle}

The total gas energy per particle is given by equation (\ref{usol}) which gives
\begin{eqnarray}\label{62}
u = U/N = \frac12 T_V
- \frac12 T_V^2 \int_0^1 dx [y'(x)]^2
\end{eqnarray}

In order to compare with the numerical simulation, the result (\ref{62})
must be adjusted in two ways:

i) we must subtract the contribution
of the constant term $-\frac16$ in the interaction energy (\ref{usol})
which was not included in the simulation;

ii) we must multiply $u$ with $m$, the particle masses, in order to
account for the fact that we rescaled both the
kinetic and interaction potential energies by one factor $m$.

We get thus the following theoretical prediction for the energy per
particle in the simulation
\begin{eqnarray}\label{62p}
u_{\rm sim} &=& \frac{U_{\rm sim}}{N} = m(u + \frac{1}{24}) =
\frac12 m T_V + \frac{1}{24} m
- \frac12 m T_V^2 \int_0^1 dx [y'(x)]^2 \\
&=& \frac12 T + \frac{M}{24N}
- \frac{1}{2M} N T^2 \int_0^1 dx [y'(x)]^2 \nonumber
\end{eqnarray}

The second term is the contribution of the constant term $-\frac16$ which
was not included in the simulation. This is
\begin{eqnarray}
\Delta U/N = \frac12 g^2 \int_0^1 dx dy \rho(x) \rho(y) \frac16 =
\frac{1}{24}\,.
\end{eqnarray}

\subsection{Critical temperature}

The critical Vlasov temperature is given by equation (\ref{Tc1}).
Taking into account the normalization factor $g^2=\frac12$ this is
$(T_V)_{c1} = \frac{1}{4\pi^2}$. Converting to the actual temperature
as $T = T_V m = T_V M/N$ we get the critical temperature
\begin{eqnarray}
T_{c1} = \frac{1}{4\pi^2 N}
\end{eqnarray}
Thus we expect to see a discontinuity in the derivative of the
caloric curve, defined as $u_{\rm sim}(T)$ with $u_{\rm sim}$
the total gas energy per particle, given in (\ref{62p}), at
\begin{eqnarray}\label{Tc}
\frac{E_{kin}}{N} = \frac12 T_{c1} = \frac{1}{8\pi^2 N}
\end{eqnarray}
We tabulated in Table~\ref{Table:2} the values of the kinetic energy
per particle at the critical temperature
$T_{c1}$ for several values of $N$ used in the simulation.

\begin{table}
\caption{\label{Table:2}
Numerical results for the kinetic energy per particle at the critical
temperature, from equation (\ref{Tc}), for the values of $N$ considered
in the simulation.}
\begin{center}
\begin{tabular}{cc}
\hline
$N$ & $\frac12 T_{c1} (\times 10^{-4})$ \\
\hline
20 & 6.33 \\
40 & 3.17 \\
80 & 1.58 \\
160 & 0.79 \\
\hline
\end{tabular}
\end{center}
\end{table}

The results of Table~\ref{Table:2} agree qualitatively with the behavior observed in Fig. \ref{fig:VaryingN_KE}---the critical temperature decreases with the number of particles. The position of
the discontinuity is reproduced reasonably well, and the agreement improves with
increasing $N$. For $N=160$ the discontinuity appears at 
$E_{\rm kin}/N=0.72\cdot 10^{-4}$ which is very close to the theory prediction 
of $0.79 \cdot 10^{-4}$.

\subsection{The caloric curve}

The simulation computed the average values of the
total gas energy $U/N$ and kinetic energy $E_{\rm kin}/N = \frac12 T$
per particle. 
We compare next the simulation result for the caloric curve with the theoretical
prediction in (\ref{62p}).

\begin{figure}[t]
\begin{center}
\includegraphics[scale=1]{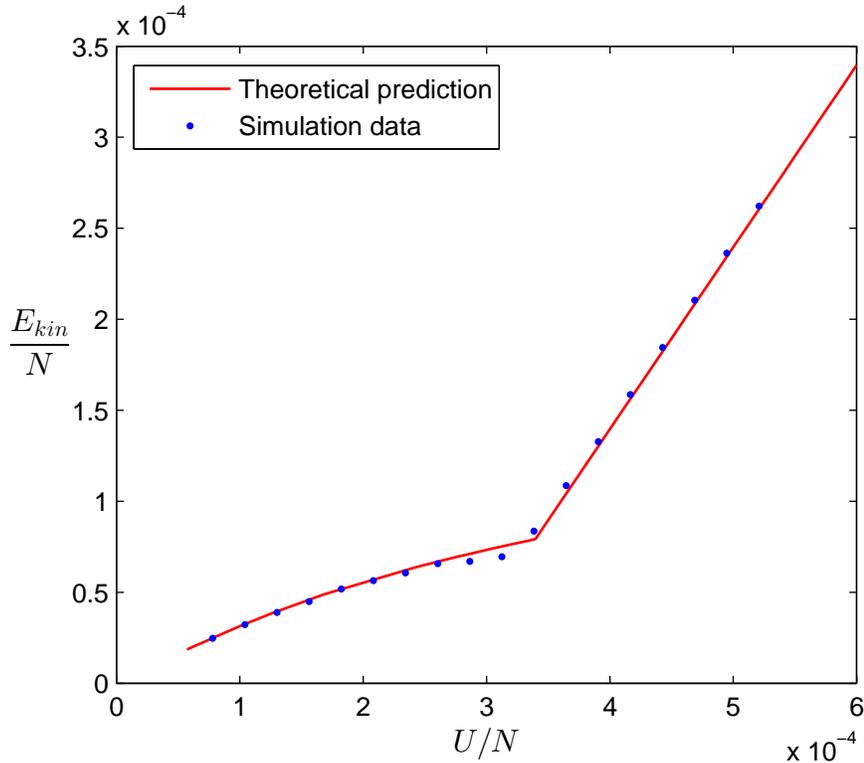}
\caption{\label{fig:N_160_TheoSim} Caloric curve: 
Theoretical prediction vs. simulation results for $N=160$. }
\end{center}
\end{figure}

Above the critical temperature the last term in the energy formula (\ref{62p})
vanishes, and we get, with the normalization of the simulation
\begin{eqnarray}
\frac{U_{\rm sim}}{N} =
\frac12 T + \frac{1}{24N} = \frac{E_{kin}}{N} + \frac{1}{24N}\,.
\end{eqnarray}
The kinetic energy per particle expressed as function of total energy
per particle is a straight line with intercept $-\frac{1}{24N}$.
For $N=160$ this intercept is $-2.604\cdot 10^{-4}$, which is very close
to the intercept of the straight line observed in Fig.~\ref{fig:N_160_with_fit_curves}. Figure \ref{fig:N_160_TheoSim} presents a comparative plot of the theoretical and simulation data graphed together.

\begin{figure}[t]
    \centering
   \includegraphics[width=5in]{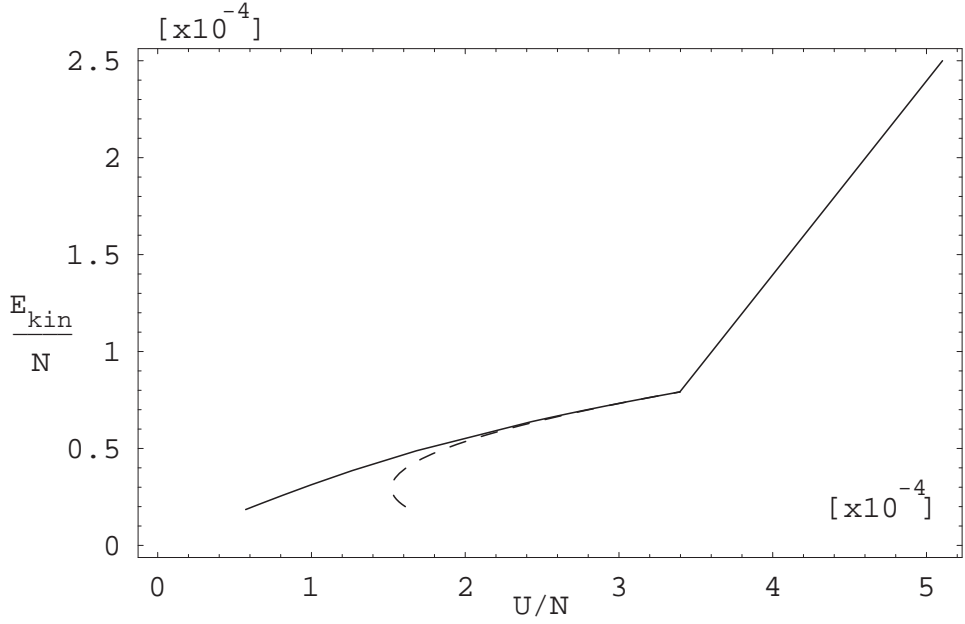}
    \caption{
Plot of the caloric curve $(U_{\rm sim}/N, E_{kin}/N)$ for $N=160$.
Dashed curve: approximative
result using the approximation $K_1(\alpha) = \pi \alpha y_0^2$.}
\label{Fig:N160}
 \end{figure}

For temperatures below the critical temperature $T < T_{c1}$
the last term in the energy formula (\ref{62p}) starts contributing. 
In this region we have
\begin{eqnarray}\label{Upp}
\frac{U_{\rm sim}}{N} &=&
- \frac12 N T^2 \int_0^1 dx [y'(x)]^2 + \frac12 T + \frac{1}{24N}\\
&=& - 2N \left(\frac{T}{2}\right)^2 K_1(\alpha) + \frac12 T + \frac{1}{24N}
\nonumber
\end{eqnarray}

An analytical approximation for the integral $K_1(\alpha)=\int_0^1 dx [y'(x)]^2$
which is valid
very close to the critical point is given in equation (\ref{Kapp})
\begin{eqnarray}
K_1(\alpha) &\simeq & \pi \alpha y_0^2 = 24\pi \alpha 
\left(\sqrt{\frac{T_{c1}}{T}} - 1\right) \\
&=& 24\pi \frac{1}{\sqrt{NT}}
\left( \sqrt{\frac{1}{8\pi^2 N}\cdot \frac{2}{T}} - 1 \right) \nonumber
\end{eqnarray}
The dashed curve in Figure~\ref{Fig:N160} represents the result for
the caloric curve following from this approximation. The solid curve shows
the exact caloric curve obtained using the exact (numerical) result for
$K_1(\alpha)$ in Table~\ref{Table:1}. This table contains a tabulation of the
integral $K_1(\alpha)$ for values of $\alpha$ from $2\pi$ to 13.
Each of these points corresponds to a value of the temperature, according to
\begin{eqnarray}
\alpha^2 = (2\pi)^2 \frac{T_{c1}}{T}
\end{eqnarray}
From this we get the kinetic energy per particle
\begin{eqnarray}
\frac{E_{kin}}{N}= \frac12 T = \frac{(2\pi)^2 T_{c1}}{\alpha^2}
\end{eqnarray}
The corresponding result for the total energy per particle is obtained
from (\ref{Upp}).
Thus for each value of $\alpha$ in Table~\ref{Table:1} we get a point with
coordinates
$(U_{\rm sim}/N, E_{\rm kin}/N)$. The set of all these points forms the caloric
curve for temperatures below the critical temperatures shown in
Figure~\ref{Fig:N160}.

\subsection{Equation of state}

The comparison of the simulation results for the pressure with the theoretical
calculation is more difficult. It is known \cite{Choquard} that for systems
with long-range interactions one has to distinguish between the thermodynamical 
pressure (computed
as $- (\partial F/\partial L)_{N,T}$) and the kinetic pressure (computed as in the
numerical simulation). Additional complications have to be taken into account
when using periodic boundary conditions \cite{Louwerse}.

In order to illustrate these
issues, consider the case of a one-dimensional system of particles interacting
by a constant attractive potential $V(x,y) = - c L$, proportional to the
volume of the system $L$ and $c>0$. (Such a constant term is present in the
interaction potential (\ref{Vbarsol}), where it appears because of  imposing
periodic boundary conditions.) The free energy is
\begin{eqnarray}
F = - cNL - k_B T_V N \log L + F_{\rm kin}
\end{eqnarray}
which yields the thermodynamical pressure
\begin{eqnarray}
p = k_B T_V \frac{N}{L} + cN
\end{eqnarray}
This is the ideal gas law, supplemented by the addition of a positive term.
On the other hand it is clear that the kinetic pressure will not be changed 
by the
constant potential $V(x,y)=-cL$ which corresponds to zero forces. This simple
argument illustrates the difficulties encountered with the
interpretation of the thermodynamical pressure in systems with
long-range interactions. For these reasons we show only the results for the
kinetic pressure obtained from the numerical simulation, see
Figure~\ref{fig:N_160_Pressure_fit_curves}.

\section{Conclusions}

We studied in this paper the thermodynamical properties of a one-dimensional
gas of $N$ particles interacting via one-dimensional gravitational potentials, 
subject to periodic boundary conditions. This results in a modification of the 
two-body interaction potential which takes into account the contributions of 
an infinite number of mirror images (Ewald sum).
The method of derivation is an application of Kiessling's approach to an 
infinite gravitational system where the potential is regularized by an 
exponential damping factor that is finally taken to the limit where the 
damping factor vanishes \cite{Kiessling}. This model was proposed in 
Ref.~\cite{MR} and was also used in Ref.~\cite{KM} to describe a plasma 
consisting of charged particles in a uniform charge background. 
In this formulation each particle carries with it a uniformly distributed 
negative mass (or charge) background that arises from its infinite replicas. 
This should not be not confused with an external background potential that 
is introduced in an ad-hoc approach. The system possesses complete 
translational invariance without imposing any additional constraints.

In carrying out our computations of the thermodynamic properties we considered the Vlasov limit, which corresponds to taking the particle number very large, at fixed volume (length) 
and total mass. In this limit the total energy and entropy
have usual extensive properties, and we derived the exact solutions for the
thermodynamical properties in the canonical ensemble.
 
In common with a gravitational system with an externally imposed background 
potential, the spatially periodic system considered here also undergoes a 
phase transition at a critical temperature $T_{c1}$ \cite{Val}. Above the 
critical temperature the gas density is uniform, while below this temperature 
becomes non-uniform and has a stable unimodal density profile that is not 
fixed in position. Thus there is a continuum of solutions which differ only 
by a translation. Both the translationally invariant system considered here 
and the rigid system with externally imposed background potential exhibit 
an infinite sequence of critical points at which the system develops 
additional, unstable states. We show that only the inhomogeneous density 
state with unimodal density distribution appearing at $T_{c1}$ is stable. 
This is in contrast with the free boundary self-gravitating system that has 
been studied extensively (for reviews see Refs.\cite{Bavaud, LevinReview}). 
For that system it was shown analytically by Rybicki that no phase transition 
occurs at any energy \cite{Rybicki} in the one-dimensional gravitational system
without hard core interaction. Note that in higher dimension it is necessary 
to screen the singularity of the gravitational force to obtain a phase 
transition\cite{Miller1998, Youngkins2000, Miller2002}.

Here we showed that the equilibrium density obeys a variant of the Lane-Emden 
equation which determines the gas density up to a translation. Both approximative
and numerically-computed exact solutions for the thermodynamical properties 
were obtained and used to evaluate the internal energy and 
heat capacity as a function of temperature. A discontinuity in the slope of 
the caloric curve and corresponding discontinuity in the heat capacity, 
manifestations of a second order phase transition, were obtained.

In addition to the theoretical derivation of the thermodynamical properties, 
we carried out dynamical $N$-body simulations of the model which confirmed 
the analytically predicted features of the phase transition at the critical 
temperature $T_{c1}$. The temperature dependencies of the numerically-computed 
averages of the per-particle energy and pressure as well as the largest 
Lyapunov exponent display sudden changes in their slopes at $T \sim T_{c1}$. 
The simulations utilized efficient event-driven algorithms that employed 
exact expressions for the time evolution of the system's phase-space and 
tangent-space vectors.

The long-range nature of the interaction potential of the model considered
introduces known difficulties in the theoretical calculation of the
equation of state in the inhomogeneous density state below the critical point. 
We plan on returning to this issue in future work. 
Nonetheless the simulation tools employed here allowed for the numerical 
estimations of the thermodynamic quantities and their corresponding behavior 
with changing temperature. Moreover, the $\mu$-space distributions obtained 
in simulation confirm the existence of inhomogeneity in density below the 
critical temperature as predicted by our analytic treatment of the system.

Finally, it is worth highlighting that the discontinuity in the slope of 
the temperature dependence of the largest Lyapunov exponent displayed by 
our simulations near the critical temperature reaffirms the previously 
reported findings that suggested the applicability of the Lyapunov exponents 
as a possible indicator of phase transitions.

\newpage
\appendix
\section{Appendix: Proof of the equation (\ref{Vexact})}

We give in this Appendix further details of the calculation of the sum over 
mirror images.
This is done by writing the sum in (\ref{sum}) as
\begin{eqnarray}
&& \sum_{k = -\infty}^\infty |x - y + k L| e^{-\kappa |x-y+kL|} \\
&& = |x-y| e^{-\kappa |x-y|} + \sum_{n=-\infty}^{-1}
(y - (x+nL)) e^{-\kappa (y-nL-x)} + \sum_{n=1}^\infty
(x+nL-y) e^{-\kappa (x+nL-y)} \nonumber \\
&=& |x-y| e^{-\kappa |x-y|} + \sum_{n=1}^{\infty}
(y - (x-nL)) e^{-\kappa (y+nL-x)} + \sum_{n=1}^\infty
(x+nL-y) e^{-\kappa (x+nL-y)} \nonumber \\
&=& |x-y| e^{-\kappa |x-y|} +
\sum_{n=1}^{\infty}
nL e^{-\kappa nL} (e^{-\kappa (y-x)} + e^{\kappa (y-x)}) \nonumber \\
& & + (y-x) \sum_{n=1}^{\infty}
e^{-\kappa nL} (e^{-\kappa (y-x)} - e^{\kappa (y-x)}) \,.
\end{eqnarray}
The sums over $n$ can be evaluated in closed form
\begin{eqnarray}
&& \sum_{n=1}^{\infty}
e^{-\kappa nL} = \frac{1}{e^{\kappa L}-1} \\
&& \sum_{n=1}^{\infty}
nL e^{-\kappa nL} = L\frac{e^{\kappa L}}{(e^{\kappa L}-1)^2} \,.
\end{eqnarray}
Substituting into the sums above gives equation (\ref{Vexact}).

\section{Derivation of the free energy}

We prove here the result (\ref{Fexact}) for the configurational contribution
to the free energy per particle $f_Q[\rho]$.
The starting point is the expression
\begin{eqnarray}\label{fQ0}
f_Q = \frac12 T \int_0^1 dx \rho(x)\log\rho(x) - \frac12 \lambda - \frac12 T
- T \log L
\end{eqnarray}
which is obtained by eliminating the double integral in (\ref{fQnorm})
using the Euler-Lagrange
equation (\ref{EL}). Multiplying (\ref{EL}) with $\frac12\rho(x)$ and integrating
over $x$ we get
\begin{eqnarray}\label{Ustart}
\frac12 g^2 \int_0^1 dx dy \rho(x)\rho(y)
(|x-y| - (x-y)^2 - \frac16 ) = - \frac12 T \int_0^1 dx \rho(x)\log\rho(x) -
\frac12\lambda - \frac12 T \,.
\end{eqnarray}
Substituting this into (\ref{fQnorm}) gives (\ref{fQ0}).

We will evaluate the integral in the first term and the Lagrange multiplier,
and will show that they are given by
\begin{eqnarray}\label{Jdef}
J &:=& \int_0^1 dx \rho(x) \log\rho(x) =
\frac{3}{2\alpha^2} \int_0^1 dx [y'(x)]^2 +
(1 - e^{y(0)} + y(0) ) \\
\label{lambdadef}
\lambda &=& - \frac{g^2}{\alpha^4} \int_0^1 dx [y'(x)]^2 +
T(e^{y(0)} - y(0) - 2) \,.
\end{eqnarray}

{\bf 1.} The calculation of the integral (\ref{Jdef}).
This is done by writing it as
\begin{eqnarray}
J &=& \int_0^1 dx \rho(x) \log\rho(x) =
\int_0^1 dx (1 - \frac{1}{\alpha^2}
y''(x)) y(x) = I_1 - \frac{1}{\alpha^2} I_2
\end{eqnarray}
where we used the Lane-Emden equation $y''(x) = \alpha^2 (1 - \rho(x))$
and evaluating the resulting integrals as follows.

There are two integrals appearing here.
\begin{eqnarray}\label{I1}
I_1 := \int_0^1 dx y(x) =
\frac{1}{2\alpha^2} \int_0^1 dx [y'(x)]^2 + 1 - e^{y(0)}+y(0)
\end{eqnarray}
This follows from the relation (energy conservation for the equivalent
dynamical problem)
\begin{eqnarray}
\alpha^2 (e^{y(0)} - y(0)) = \frac12 [y'(x)]^2 + \alpha^2 (e^{y(x)} - y(x))
\end{eqnarray}
and integration over $x:(0,1)$ using the normalization condition
$\int_0^1 dx e^{y(x)} = 1$.

The second integral is
\begin{eqnarray}\label{I3}
I_2 := \int_0^1 dx y''(x) y(x) = y'(1) y(1) - y'(0) y(0) - \int_0^1 dx [y'(x)]^2
= - \int_0^1 dx [y'(x)]^2
\end{eqnarray}
where we used the boundary conditions $y(0)=y(1), y'(0)=y'(1)$.

{\bf 2.} Next we compute the Lagrange multiplier $\lambda$. This is expressed
by taking
$x=0$ in the Euler-Lagrange equation (\ref{EL}) which gives
\begin{eqnarray}
\lambda = - g^2\int_0^1 dx \rho(x) (x- x^2-\frac16) - T (\log\rho(0) + 1)
\end{eqnarray}
The integral appearing here is evaluated by integration by parts. This is
\begin{eqnarray}\label{I4}
I_3 &:=& \int_0^1 dx e^{y(x)}(x - x^2 - \frac16) =
\int_0^1 dx (1 - \frac{1}{\alpha^2} y''(x))
(x - x^2 - \frac16) \\
 &=& - \frac{1}{\alpha^2} \int_0^1 dx y''(x)
(x - x^2 - \frac16) = - \frac{1}{\alpha^2} y'(x)(x - x^2 - \frac16)|_0^1 +
\frac{1}{\alpha^2} \int_0^1 dx y'(x)(1 - 2x) \nonumber \\
 &=& \frac{1}{\alpha^2} y(x)(1-2x)|_0^1 + \frac{1}{\alpha^2} 2\int_0^1 dx y(x)
\nonumber \\
 &=& -\frac{2}{\alpha^2} y(0) + \frac{2}{\alpha^2} I_1 \nonumber
\end{eqnarray}
where the integral $I_1$ is given in (\ref{I1}).

We get finally the result for the Lagrange multiplier
\begin{eqnarray}
\lambda = - g^2 I_3 - T(y(0) + 1)
\end{eqnarray}
where the integral $I_3$ is given in (\ref{I4}). Combining all terms
gives the result (\ref{lambdadef}).

\begin{table}
\caption{\label{Table:1}
Numerical solutions of the equation (\ref{y0eq}) for $\alpha \geq 2\pi$.
Only the solution with $y_0<0$ is given; all other solutions can be obtained
from this by a translation.
We give also the values of the integral $K_k(\alpha)$
defined in (\ref{Kdef}). The column $\delta_k(\alpha) = 1-e^{y_0}+y_0+1/\alpha^2 K_k$
gives the function appearing in the free energy difference with the $k=0$ mode,
see equation (\ref{DeltaF}).}
\begin{center}
\begin{tabular}{c|ccc|ccc}
\hline
$\alpha$ & $y_1(0)$ & $K_1(y_0)$ & $\delta_1(\alpha)$
         & $y_2(0)$ & $K_2(y_0)$ & $\delta_2(\alpha)$ \\
\hline\hline
$2\pi$ & 0        & 0        & 0        & - & - & - \\
6.3    & -0.2646  &  1.27286 & -0.00004 & - & - & - \\
6.4    & -0.7495  &  9.01418 & -0.00203 & - & - & - \\
6.5    & -1.06682 &  17.0712 & -0.00687 & - & - & - \\
6.6    & -1.335   & 25.4278  & -0.01442 & - & - & - \\
6.7    & -1.5800  & 34.1811  & -0.02453 & - & - & - \\
6.8    & -1.8084  & 43.2443  & -0.03710 & - & - & - \\
6.9    & -2.0267  & 52.6788  & -0.05200 & - & - & - \\
7.0    & -2.2377  & 62.4879  & -0.06914 & - & - & - \\
8.0    & -4.2155  & 184.617  & -0.34562 & - & - & - \\
9.0    & -6.2459  & 362.172  & -0.77658 & - & - & - \\
10.0   & -8.4725  & 613.968  & -1.33303 & - & - & - \\
11.0   & -10.9388 & 960.43   & -2.00138 & - & - & - \\
12.0   & -13.6589 & 1423.37  & -2.77439 & - & - & - \\
\hline
$4\pi$ & -15.3135  & 1745.99 & -3.25689 & 0 & 0 & 0 \\
12.6   & -15.4145  & 1766.68 & -3.28651 & -0.2646 & 5.09144 & -0.00004 \\
12.7   & -15.7162  & 1829.17 & -3.37532 & -0.5508 & 20.4223 & -0.00067 \\
12.8   & -16.0205  & 1893.24 & -3.46508 & -0.7495 & 36.0567 & -0.00203 \\
12.9   & -16.3272  & 1958.87 & -3.55585 & -0.917  & 52.0220 & -0.00410 \\
13.0   & -16.6366  & 2026.14 & -3.64761 & -1.0668 & 68.2824 & -0.00687 \\
\hline
\end{tabular}
\end{center}
\end{table}

\begin{acknowledgments}
The authors benefited from insightful discussions with Dragos Anghel, Carlos Schat, Igor Prokhorenkov, Michael Kiessling, and Marc Miller.
\end{acknowledgments}

\newpage
\bibliography{PRE_1DGravGas}

\end{document}